\documentclass[12pt]{article}
\usepackage{amssymb,amsmath}
\usepackage{bm}
\usepackage{comment}
\usepackage[dvips]{graphicx}
\usepackage{cite}
\usepackage{subfigure}

\usepackage{bm}
\usepackage{booktabs}

\def\beq{\begin{equation}}
\def\eeq{\end{equation}}
\def\beqa{\begin{eqnarray}}
\def\eeqa{\end{eqnarray}}
\def\d{\partial}

\def\bra{\langle}
\def \ket{\rangle}
\def\Lam{\Lambda}

\newcommand{\1}{\mbox{1}\hspace{-0.35em}\mbox{1}}

\begin{document}
\baselineskip 0.7cm

\begin{titlepage}

\begin{flushright}
{\bf KUNS-2343 }\\
\today
\end{flushright}

\vskip 1.35cm
\begin{center}
{\bf{Solving the Naturalness Problem by Baby Universes in the Lorentzian Multiverse}}

\vskip 1.2cm
Hikaru Kawai\footnote{hkawai@gauge.scphys.kyoto-u.ac.jp}, Takashi Okada\footnote{okada@gauge.scphys.kyoto-u.ac.jp},
\vskip 0.4cm

{\em{ Department of Physics, Kyoto university, Kyoto 606-8502, Japan}}\\

\vskip 1.5cm

\abstract{
We propose a solution of the naturalness problem  in the context of the multiverse wavefunction $without$ the anthropic argument. 
If we include microscopic wormhole configurations in the path integral,	the wave function becomes a superposition of universes with various values of the coupling constants such as  the cosmological constant, the parameters in the Higgs potential, and so on. We analyze the quantum state of the multiverse, and evaluate the density matrix of one universe. We show that the coupling constants induced by the wormholes are fixed in such a way that the density matrix is maximized. In particular, the cosmological constant, which is in general time-dependent, is chosen such that it takes an extremely small value in the far future. We also discuss the gauge hierarchy problem and the strong CP problem in this context. Our study predicts that the Higgs mass is $m_h = 140 \pm 20$ GeV and $\theta=0$. }
\end{center}

\end{titlepage}

\section{Introduction and Conclusion}
One of the major problems of particle physics  and cosmology is the smallness of the observed value of the vacuum energy, that is the cosmological constant $\Lam$. We must explain why $\Lam$ is many orders of magnitude smaller than the  Planck scale\cite{wormhole}. One of the most promising attempts to solve this problem is the one based on the Euclidean wormhole effect first proposed by Coleman \cite{coleman1988there}\footnote{In \cite{banks1988prolegomena}, Banks also discussed the effect of bi-local interaction. In this paper, we mainly follow Coleman's argument.}and studied closely by other authors\cite{klebanov1989wormholes, polchinski1989phase, giddings1989baby,PhysRevLett.62.1429,grinstein1988light,preskill1989wormholes,cline1989can,rubakov1988third,strominger1989lorentzian,fischler1989quantum,cline1989does}. 
In this paper, we discuss the wormhole effect in the context of the Lorentzian multiverse\footnote{Although \cite{strominger1989lorentzian,fischler1989quantum,cline1989does} also studied the wormhole effect in the Lorentzian gravity, 
our mechanism is different from the previous work as we will discuss in Section \ref{sec:comparison}.}, and 
propose a mechanism to solve the naturalness problems such as the cosmological constant, the gauge hierarchy, and the strong CP problem.

To explain the motivation of this paper, we begin by briefly discussing Coleman's solution to the cosmological constant problem (see Section \ref{sec:baby} for the details of the derivation of the following equations).  
We start with the path integral of the Euclidean gravity. If we take microscopic wormhole configurations into account, the following interaction $\Delta S$ is induced in addition to the original action;
\begin{equation}
\Delta S= \sum _{i} (a_i +a_i^\dagger)C_i\int d^4 x \sqrt{g}\mathcal{O}_i ,\label{eq:babyeffect}
\end{equation} 
where $a_i,\ a_i^\dagger$ are the annihilation and creation operators of the type $i$ babyuniverse. Then, the partition function of the parent universe is given by an integral over the eigenvalues of $a_i+a_i^\dagger$.  

\begin{figure}[htbp]
\begin{center}
\includegraphics[width=5cm]{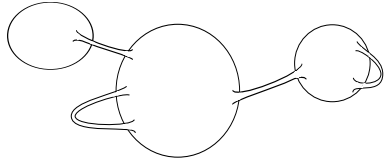}
\caption{A sketch of an example of the Euclidean multiverse. Parent universes are interacting through baby universes.}
\label{fig:euclidean_multiverse}
\end{center}
\end{figure}

For example, if we focus on the identity operator $\mathcal{O}=1$, the partition function becomes \begin{equation*}
Z_{\text{universe}} = \int \mathcal{D}g d\Lam \exp\bigl({-\int d^4x \sqrt{g}(R+2 \Lam)} \bigr),
\end{equation*}
where the wormhole effect results in the integration over $\Lambda$. The path integral over the metric $g$ can be approximated by a 4-sphere solution, whose action is proportional to $\frac{1}{\Lam}$. Therefore we have
\begin{equation*}
Z_{\text{universe}} \sim \int d\ {\Lam} \ e^{\frac{1}{\Lam}},
\end{equation*}
and the integrand has a strong peak at $\Lam \sim 0$.
Furthermore, if we consider the multiverse, in which universes are connected each other through  baby universes (see Fig.\ref{fig:euclidean_multiverse}), the above integrand is replaced to $\exp(\exp(\frac{1}{\Lam}))$, and the peak gets stronger.
Based on this argument, Coleman claimed that the cosmological constant problem could be solved by the wormholes.\\

\begin{figure}[htbp]
\begin{center}
\includegraphics[width=7cm]{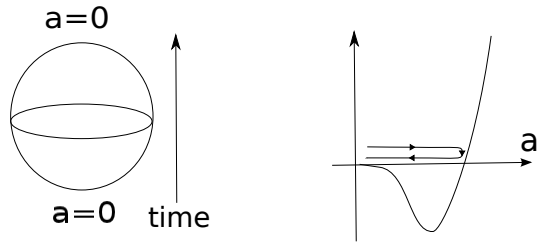}
\caption{The 4-sphere solution can be interpreted as a foliation of 3-spheres whose radius expands from zero to $\frac{1}{\sqrt{\Lam}}$ and then shrinks to zero.}
\label{fig:bounce}
\end{center}
\end{figure}

What does this argument imply to  Lorentzian spacetime? Naively, the 4-sphere solution is interpreted as a bounce solution. Therefore, the exponential of the action, $e^{\frac{1}{\Lam}}$, is expected to give the amplitude of a universe tunneling form nothing to the size of the 4-sphere (see Fig.\ref{fig:bounce}). 
However, if we computes the tunneling amplitude directly by the WKB method as Vilenkin did\cite{vilenkin1984quantum}, we obtain a factor $e^{-\frac{1}{\Lam}}$, instead of $e^{\frac{1}{\Lam}}$. In this sense, the physical meaning of the 4-sphere solution is not clear, and whether or not  Coleman's mechanism works in the physical Lorentzian spacetime is doubtful.

In this paper, in order to clarify this point, we study the wave function of the Lorentzian multiverse consisting of infinitely many parent universes which are interacting with each other via wormholes\cite{kawai2011asymptotically}. We will show that the density matrix of one universe has a strong peak in the space of the coupling constants induced by the wormholes. This indicates that ``the big fix" indeed occurs, that is, the coupling constants are determined dynamically by the quantum gravity. In particular, the multiverse wave function predicts that the cosmological constant in the far future becomes extremely small. We  will also find that the wormhole effect  fixes the other coupling constants such as the Higgs parameters and the strong CP  phase in the standard model.

This paper is organized as follows.  In Section \ref{sec:baby}, we review the derivation of the effective action \eqref{eq:babyeffect} and obtain its Lorentzian counterpart via a Wick rotation (see Fig.\ref{fig:multiverse}, which is the Lorentzian version of Fig.\ref{fig:euclidean_multiverse}). We see that because of the wormholes the wave function of the parent universes becomes a superposition of states with various values of the coupling constants $\{ \lambda_i\}$.

In  Section \ref{sec:universe}, for the fixed coupling constants $\{ \lambda_i\}$, we calculate the wave function of a parent universe $\phi_{E=0}(z)$, where $z$ is the size\footnote{Strictly speaking, $z \equiv  a^3/9$ has a dimension of volume. However, for the sake of simplicity, we call it ``size''.} of the universe, by using the WKB approximation. We assume that the parent universes have the topology of $S^3$, and use  the superminispace approximation for each of them
\begin{equation}
ds^2 =\sigma^2( - N(t)^2 dt^2 + a(t)^2 d{\Omega}_3^2), \label{eq:metric}
\end{equation}
where $d{\Omega}_3^2$ is the metric of unit $S^3$. We also assume that they are created from nothing at a small size $\epsilon$ via some tunneling process. Then the wave function of each parent universe is given by
\begin{eqnarray}\phi_{E=0}(z) = \frac{1}{\sqrt{\pi/2}\sqrt{z}\sqrt{k_{E=0}(z)}}\sin (\int_{0}^z k_{E=0}(z')dz' +\alpha )\label{eq:wkb_intro},
\end{eqnarray}
where $E=0$ represents the so-called Hamiltonian constraint, which we will discuss later, and $k_{E=0}$ is defined by
\begin{eqnarray}
k_{E=0}^2(z) &\equiv& -2U(z) \\
&=&9\Lambda
-\frac{9^{1/3}}{z^{2/3}}K
+\frac{2M_{matt}}{z}+\frac{2S_{rad}}{z^{4/3}}
-\frac{2E}{z} \label{eq:potential_intro}.
\end{eqnarray}
Here $\Lambda, M_{matt}, S_{rad}$ are the cosmological constant, the amounts of matter and radiation, respectively. In principle, they are determined by examining the time evolution of the universe, once its initial condition at $z=\epsilon$ and the coupling constants $\{\lambda_i \}$ are given. In this sense, they depend on the coupling constants $\{\lambda_i \}$ as well as on time, or $z$.
\beq
\Lambda=\Lambda(\{\lambda_i \},z),\ M_{matt}= M_{matt}(\{\lambda_i \},z),\ S_{rad}=S_{rad}(\{\lambda_i \},z).
\eeq
For example, if some matter decays into radiation at some $z$, $S_{rad}$  increases at this point. 
The factor $\frac{1}{\sqrt{k_{E=0}(z)}}$ in \eqref{eq:wkb_intro} behaves like $\frac{1}{\Lam^{1/4}}$ for large $z$ and plays an important role for our mechanism. 

In Section \ref{sec:multiverse}, we construct the wave function of the multiverse. The N-universe state $| \Phi_N \ket$ is obtained by taking a tensor product of N universes and superposing over $\{ \lambda_i \}$:
\begin{equation}
\Phi_N(z_1,\cdots, z_N)\sim
\int d \vec{\lambda} \ \mu^{N}\phi_{E=0}(z_1)\cdots\phi_{E=0}(z_N)\otimes \ 
w(\vec{\lambda})
 |  \vec{\lambda}\ket,\label{eq:z_rep}
\end{equation}
where $\vec{\lambda}$ represents the set of induced coupling constants $\{\lambda_i\}$.  $| \vec{\lambda}\ket$ is the eigenstate of $a_i+a_i^\dagger$ with eigenvalue $\{\lambda_i\}$, and $w(\vec{\lambda})$ is the initial wave function of the baby universes.
$\mu$ is the probability amplitude of creating one universe. 
Then, the multiverse state can be written as 
\beq
| \phi_{multi} \ket = \sum_{N=0}^\infty |\Phi_N \ket,\label{eq:N_intro}
\eeq
where $|\Phi_N \ket$ is the N-universe state whose $z$-representation is given by \eqref{eq:z_rep}.
Then the density matrix of our universe is obtained by tracing out the other universes. The summation over $N$ results in an exponential, and we have
\begin{eqnarray}
\begin{split}
\rho(z',z) \propto \int_{-\infty}^{\infty}  d \vec{\lambda} \  |w(\vec{\lambda})|^2|\mu|^2\ \phi_{E=0}(z')^* \phi_{E=0}(z)\times \exp\biggl(\int dz^{''} |\mu \phi_{E=0}(z^{''})|^2 \biggr). \label{eq:densitymat_intro} 
\end{split}
\end{eqnarray}



 In Section \ref{sec:vanishing_cc}, we try to fix the cosmological constant $\Lambda$ by examining the $\Lambda$ dependence of the above density matrix. If $\Lambda<0$, $\phi_{E=0}$ is exponentially suppressed for large $z$, and the exponent in the RHS of the density matrix \eqref{eq:densitymat_intro} takes some finite value. On the other hand, if $\Lambda \geq0$, it is calculated as
 \begin{eqnarray}
|\mu|^2 \int dz'' \ \frac{1}{z'' k_{E=0}}&\sim &|\mu|^2 \int dz'' \frac{1}{z''}\frac{1}{\sqrt{9\Lambda}} \label{eq:intro_exponent},
 \end{eqnarray}
because the leading behavior of the momentum for large $z$ is given by $k^2_{E=0}= 9\Lambda+\cdots$.  Since this integral is logarithmically divergent, we introduce an infrared cutoff $z_{IR}$ for $z$, so that the above integral becomes 
 \beq
 |\mu|^2 \frac{1}{\sqrt{9\Lambda}} \log z_{IR}\label{eq:intro_exponent2}. 
  \eeq
Thus we find that the integrand of \eqref{eq:densitymat_intro} has an infinitely strong peak at $\Lambda=0$, which means that the cosmological constant in the far future is automatically tuned to zero. Although we can not specify the origin of $z_{IR}$ at this stage, it is natural to consider that a sort of infrared cutoff should appear in any constructive definitions of quantum gravity. For example, in the dynamical triangulation\cite{Ambjorn:2005qt}, the number of simplexes corresponds to the infrared cutoff, or in matrix models, in which space-times emerge dynamically, it is provided by the size of the matrices.
 
However, there is a subtlety here. There is a critical value of $\Lambda=\Lambda_{cr}$ such that for $\Lambda<\Lambda_{cr}$, a classically forbidden region, $k^2_{E=0}<0$, appears in $z$-space (see Figure \ref{fig:closed_series}), and a tunneling suppression factor should be multiplied to (\ref{eq:intro_exponent2}).
Thus, for fixed $S_{rad}$ and $M_{matt}$, the density matrix becomes maximum when $ \Lambda {=}\Lambda_{cr}$.
For example, if we assume the radiation dominated universe and set $M_{matt}=0$, we have $\Lambda_{cr}= 1/S_{rad}$, and
the cosmological constant is fixed at 
   \beq
    \Lambda {=}\Lambda_{cr}=1/S_{rad}.
   \eeq
Once it is done,
 \eqref{eq:intro_exponent} becomes
 \begin{eqnarray}
|\mu|^2 \int dz'' \ \frac{1}{z k_{E=0}}\propto |\mu|^2 \sqrt{S_{rad}}  \log z_{IR} \label{eq:intro_max_rad}.
 \end{eqnarray}
  Recalling that $S_{rad}$ also depends on the induced coupling constants $\{ \lambda_i \}$, the above equation shows that  $\{\lambda_i \}$ are fixed at the values where $S_{rad}$ becomes maximum. Therefore, the value of $\Lambda$ is given by 
 \beq
\Lambda \simeq 1/ \underset{ \vec{{\lambda}}  }{\text{max}}\ S_{rad}(\vec{\lambda} ).
 \eeq
Since $S_{rad}$ is proportional to the volume of the universe, if the universe is sufficiently large, $S_{rad}$ is large and $\Lambda$ is close to zero. 

To summarize, the wormhole effect makes the wave function of the multiverse a superposition of various values of coupling constants, but they are fixed in such a way that the radiation in the far future is maximized. We call it the $big\  fix$ following Coleman. In particular, the cosmological constant is fixed as its value in the far future becomes almost vanishing.

We can give an intuitive interpretation of the above mechanism. 
The exponent in the density matrix \eqref{eq:densitymat_intro} turns out to be the time that it takes for the universe to expand from the size $\epsilon$ to $z_{IR}$. To see this we rewrite it as
\beq
\int {dz}\ |\phi_{E=0} (z)|^2 =  \int^{z_{IR}}_{\epsilon}{dz}\ \frac{1}{z k_{E=0}(z)}= \int dt  \label{eq:intro_lifetime_wkb},
\eeq 
where we have used the classical relation $k\sim \dot{z}/z$. Thus, the exponent is nothing but the $lifetime$ of the universe. Naively, smaller $\Lambda$ is favored because then the universe expands slowly (see Figure \ref{fig:classical_shape}). However, for $\Lambda<\Lambda_{cr}$, the universe bounces back to a small size in a finite time. Therefore, the $lifetime$ of the universe becomes maximum when $\Lambda=\Lambda_{cr}$. We note that the enhancement arises from the large $z$ region $z\sim z_{IR}$, where the universe can be described by the classical mechanics, which justifies treating the matter and radiation classically as in \eqref{eq:potential_intro}.  On the other hand, the quantum mechanical nature of the wormholes reflects in superposing the states with various $\{ \lambda_i \}$. In Section \ref{sec:comparison}, we compare our mechanism of the big fix with the previous works by other authors.

In Section \ref{sec:bigfix}, as an illustration of the big fix, we consider the parameters in the Higgs sector in the standard model, that is, the VEV $v_h$ and the quartic coupling constant $\lambda_h$.  We assume that the other coupling constants are fixed to their observed values.  We consider the case that $S_{rad}$ in the far future consists of the decay products of protons. Then, we can show that $S_{rad}$ is maximized when $N_b^2 m_p^{2} \tau_p$ is maximized (see around \eqref{eq:decay_conservation}), where $N_b$, $m_p$ and $\tau_p$ are the total baryon number before the decay, the proton mass and the proton lifetime, respectively. Naively, this  seems to be maximized when $m_p= m_p(v_h)$ is minimized because in the usual GUT we have
\beq
\tau_p \propto m_p^{-5}.
\eeq
Then, the wormhole mechanism seems to select out $v_h =0$ because the proton mass  $m_p$ depends on $v_h$ monotonically as follows
\begin{eqnarray}
m_p (v_h) =   M_p^{(0)} + 3 \times m_{u,d} (v_h),
\end{eqnarray}
where  $ M_p^{(0)}$ is the proton mass in the chiral limit, and $m_{u,d}$ is the current quark mass, which is proportional to $v_h$. 
However, the mass of the decay products also depends on $v_h$, and as we will show, it is in fact possible that $m_p^{2} \tau_p $ becomes minimum at some nonzero value of $m_{u,d} (v_h)$. 

Assuming that the Higgs VEV is fixed at the observed value, i.e. 246 GeV, we next consider the Higgs mass.  $\lambda_h$-dependence enters into the above combination $N_b^2 m_p^{2} \tau_p$  through the sphaleron process if we assume the leptogenesis.  Smaller $\lambda_h$ makes the sphaleron process happen more frequently and produces more baryons $N_b$. Combining this with the fact that the stability of the potential requires a lower bound on $\lambda_h$,  we can deduce that the smallest possible value of $\lambda_h$ is chosen by the big fix. This means that the physical Higgs mass should be at its lower bound, that is, around $140\pm20$ GeV \cite{holland2005triviality}. 

We then consider the strong CP problem. We analyze how the combination $N_b^2 m_p^{2} \tau_p$ depends on $\theta_{QCD}$, and find that it becomes maximum at $\theta_{QCD}=0$, which means $\theta_{QCD}$ is fixed to zero by the big fix.

In Section \ref{sec:others}, we study universes with other topologies than $S^3$. So far, we have assumed that all the parent universe have the topology of $S^3$. If we allow universes with various topologies to emerge, we must sum over them in the multiverse wave function.  Then, the density matrix 
is modified to 
 \begin{eqnarray}
\begin{split}
\rho(z', z) \propto \int_{-\infty}^{\infty}  d \vec{\lambda}\  w(\vec{\lambda})^2|\mu|^2\ \phi_{E=0}(z')^* \phi_{E=0}(z)\times \exp\biggl( \sum_{\alpha''}\int dz^{''} |\mu_{\alpha''} \phi^{(\alpha'')}_{E=0}(z^{''})|^2 \biggr),  \label{eq:dens_gen_intro}
\end{split}
\end{eqnarray}
where $\alpha$ labels the topology of the universe, and $\mu_\alpha$ is the probability amplitude to create such universe. Thus, the exponent of the density matrix is the sum of contributions from various topologies. We repeat the same analysis as $S^3$ for the other topologies, and find that the flat universes ($K=0$)\footnote{Since we assume the universes are spatially compact, the topology of  flat universe is actually torus.} have the largest contribution. In this case, the vanishing of the asymptotic cosmological constant is still valid, while the analysis of the big fix is modified rather drastically.


\section{Effect of Baby Universes} \label{sec:baby}

We first review Coleman's argument on the effect of the baby universes \cite{coleman1988there}(see also \cite{klebanov1989wormholes}). We start from the Euclidean Einstein gravity with a bare cosmological constant  $\Lambda_0$,
\begin{equation}
\int\mathcal{D} g \exp( - S_E)=\int\mathcal{D} g_{\mu \nu} \exp(-\int d^4x \sqrt{g}(R+ 2 \Lambda_0)). \nonumber
\end{equation}

\begin{figure}[btm]
\begin{center}
\includegraphics[width=3cm]{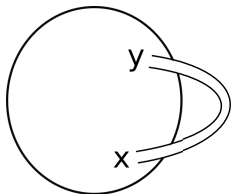}
\caption{A wormhole induces bi-local interactions at its legs.}
\label{fig:bilocal}
\end{center}
\end{figure}

A Planck-size wormhole configuration effectively adds to the partition function the following bi-local interactions (see Figure.\ref{fig:bilocal}),

\begin{equation}
\int \mathcal{D}g\ \frac{1}{2} 
\ c_{ij} e^{-2S_{\text{wh}}} \int d^4 x d^4 y
\sqrt{g(x)}\sqrt{g(y)}
O^i(x)O^j(y)  \exp(-S_E),
\end{equation}
where  the repeated indices $i, j$ are contracted. $c_{ij} $  are some constants, and $2S_{\text{wh}}$ is the action of the wormhole. 
Summing over the number of wormholes amounts to the factor
\begin{equation}
\exp\biggl(\frac{1}{2} 
e^{-2S_{\text{wh}}} \int d^4 x d^4 y \nonumber
\sqrt{g(x)}\sqrt{g(y)}
O^i(x)O^j(y) \biggr).
\end{equation}
By introducing auxiliary variables $\lambda_i$, the bi-local interactions can be rewritten as local interactions as follows,
\begin{equation}
\int \biggl[ \prod_i d\lambda_i \biggr] \exp\biggl(- e^{-S_{wh}} \lambda_i \ \int d^4 x \sqrt{g(x)} O^i(x)- \frac{1}{2}\lambda_i d^{ij} \lambda_j \biggr),\label{eq:general_int}
\end{equation}
where $d^{ij}$ is the inverse of the matrix $c_{ij}$. For example, the identity operator  $O^{1}(x)= \hat{1}$ ($i=1$) shifts the $bare$ cosmological constant  $\Lambda_0$ linearly; $\Lambda_0 \rightarrow \Lambda_0 + e^{-S_{wh}}\lambda_{\hat{1}}$, which becomes a variable to be integrated over.

Alternatively, we can express the  wormhole effect by  using  the following  Lagrangian
\begin{equation}
S_{\text{eff}}=S_E + e^{-S_{\text{wh}}}\sum_i (a_i^\dagger+a_i)\int d^4 x\sqrt{g
(x)} O^i(x), \label{eq:creationannihilation}
\end{equation}
 where we have introduced  pairs of operators  $a_i$ and  $a_i^\dagger$ satisfying $[a_i,a_j^\dagger]=c_{ij}$, which can be interpreted as the creation/annihilation operators of the baby universe of type $i$. To understand this formula, one considers an amplitude between the initial and final state both with no baby universe $\bra\Omega | \exp\bigl(e^{-S_{\text{wh}}} \sum_i (a_i+a_i^\dagger) \int d^4x \sqrt{g} O^i \bigr) |\Omega \ket $. 
By using the Baker-Campbell-Hausdorff formula, it is easy to show that this amplitude recovers  Eqn.(\ref{eq:general_int}).
 Although  (\ref{eq:general_int}) and  (\ref{eq:creationannihilation}) are equivalent,  (\ref{eq:creationannihilation})  is more convenient  to construct the wave function of the universe.

Finally, we obtain the Lorentzian counterpart by the Wick rotation,
\begin{equation}
S=\int d^4 x \sqrt{-g(x)}(R- 2 \Lambda_0)-e^{-S_{\text{wh}}} \sum_i (a_i^\dagger+a_i)\int d^4 x\sqrt{-g(x)}O_i(x). \label{eq:start}
\end{equation}
 We use this action to study the naturalness problem.

\section{Wave Function of the Universe}\label{sec:universe}
 In this section, we forget about the wormhole effect for a while, and consider the wave function of a parent universe for the fixed coupling constants $\lambda_i$. We quantize the system of the mini-superspace via path integral, and determine the wave function by the WKB method. However, as we will discuss in section \ref{sec:vanishing_cc}, the whole picture about the big fix does not depend on these approximations, but holds quite generally.

\subsection{Wave Function of a Parent Universe}\label{eq:parent_wavefunction}

We start from the Einstein-Hilbert action,
\begin{equation*}
\int\mathcal{D} g_{\mu \nu}  \exp(i S_{\Lambda})=\int\mathcal{D} g_{\mu \nu}  \exp(i\frac{1}{16\pi G} \int d^4 x \sqrt{-g(x)}(R- 2 \Lambda)).
\end{equation*}
We will consider the homogeneous, isotropic and spatially compact universe:
\begin{equation}
ds^2 =\sigma^2( - N(t)^2 dt^2 + a(t)^2 ds_{spatial}^2), \label{eq:metric}
\end{equation}
where $\sigma^2=\frac{2G}{3\pi}$, and $ds_{spatial}^2$ is the metric on the spatial hypersurface, which has a constant curvature $K_\alpha=1,0,-1$, depending on its topology $\alpha$.\footnote{The spatial topology of the universe is torus and sphere for $K_\alpha=0,-1$ respectively. However, there are many  topologies for $K_\alpha=-1$.}

Substituting the metric (\ref{eq:metric}), the action becomes
\begin{eqnarray*}
S_{\Lambda} &=& -\frac{1}{2} \int dt \  N\bigl[a \dot{a}^2/N^2-(K_\alpha a-\Lambda a^3)\bigr],
\end{eqnarray*}
where we have written $\frac{2 G \Lambda}{9\pi}$ by the same symbol $``\Lambda"$, which is the $dimensionless$ cosmological constant. 
In terms of $z(t) := \frac{a(t)^3}{9}$, it can be expressed as
\begin{equation*}
S_\Lam= -\frac{1}{2} \int dt \  N\bigl[ \dot{z}^2/z N^2-\bigl(K_\alpha (9z)^{1/3}- 9 \Lambda z \bigr)\bigr].
\end{equation*}
The momentum $p_z$ conjugate to $z$ is given by $p_z = -\dot{z}/zN$, and
the Lagrangian can be rewritten in the canonical form,
  \begin{equation*}
L_\Lam = p_z \dot{z} - N \mathcal{H}_\Lam,
\end{equation*}
where
\begin{eqnarray}
\mathcal{H}_\Lambda(p_z,z) 
&:=& z[-\frac{1}{2}p_z^2- U(z)] ,\ \ 
 \text{where\  } U(z) :=\frac{9^{1/3}}{2 z^{2/3}}K_\alpha-\frac{9}{2}\Lambda  \label{eq:ham_kari} .
\end{eqnarray}

To describe a more realistic universe, we need to consider the energy densities of various fields. Then, instead of  \eqref{eq:ham_kari}, the potential $U$ is replaced to 
\begin{eqnarray}
\mathcal{H}_\Lambda(p_z,z) 
&:=& z[-\frac{1}{2}p_z^2- U(z)] ,\ \ \nonumber \\
\text{with  } 
&&2U(z)=-9\Lambda
+\frac{9^{1/3}}{z^{2/3}}K_\alpha
-\frac{2M_{matt}}{z}-\frac{2S_{rad}}{z^{4/3}},\label{eq:ham}
\end{eqnarray}
where the last two terms represent the radiation and matter energy ,respectively, and the associated powers of $z$  are determined by their scaling behavior, $\rho_{matt.} \propto a(t)^{-3}$ and $\rho_{rad.}\propto a(t)^{-4}$. We note that the coefficients depend on $z$ and $\lambda_i$;
\beq
\Lambda=\Lambda(\{\lambda_i \};z),\ M_{matt}= M_{matt}(\{\lambda_i \};z),\ S_{rad}=S_{rad}(\{\lambda_i \};z).
\eeq
In principle, they can be determined by solving the time evolution of the theory with coupling constants $\{\lambda_i \}$, if the initial condition of the universe is completely specified.  For example, $\Lambda$ changes its value at the end of  the inflation, and a portion of $M_{matt}$ may convert to $S_{rad}$ when some matter decays into radiation. 

To quantize this system via path integral, we take the following metric on the configuration space 
\begin{equation}
|| \delta g_{\mu \nu }||^2= \int d^4 x \sqrt{-g}g_{\mu \nu} g_{\rho \lambda} \delta g^{\mu \rho}\delta g^{\nu \lambda} \propto \int dt ( \frac{a^3}{N} (\delta N)^2+ N a (\delta a)^2),
\end{equation}
which is invariant under the general coordinate transformation, and  leads to the volume form of the functional integral
\begin{equation}
\underset{t}{\Pi}  \ a^2 \delta N \delta a  \propto \underset{t}{\Pi}  \delta N \delta z := [dN][dz].
\end{equation}
Collecting these results,  we find that the universe is described by the following path integral,
\begin{eqnarray}
\int [dN][dz][dp_z] \exp(i \int dt (p_z \dot{z} - N \mathcal{H}_\Lambda)),
\end{eqnarray}
where $\mathcal{H}_\Lambda$ is given by (\ref{eq:ham}).

\begin{figure}[bthm]
\begin{center}
\includegraphics[width=4cm]{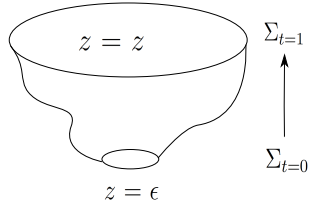}
\caption{The path integral (\ref{eq:pathintegral}) is defined as a sum over all histories connecting two geometries.}
\label{fig:history}
\end{center}
\end{figure}

In the rest of this section,
we will determine the wave function of the universe, assuming that  it initially has a small size $\epsilon$ (see Fig\ref{fig:history}), 
The amplitude between $z=\epsilon$ and $z=z$ is given by the following path integral\footnote{This analysis is similar to that of \cite{cline1989does}.},
\begin{eqnarray}
\bra z |e^{-i \hat{\text{H}}} |\ \epsilon \ket=\underset{z(0)=\epsilon,\ z(1)=z}{\int} [dp_z][dz][dN]\exp(i \int_{t=0}^{t=1}dt\  (p_z \dot{z} -N(t) \mathcal{H}_{\Lam})). \label{eq:pathintegral}
\end{eqnarray}
By choosing the gauge such that $N(t)$ is a constant $\text{T}$,  the path integral of $N(t)$ is reduced to the ordinary integral over $-\infty<\text{T}<\infty$\footnote{To be precise, we should integrate only positive $T$ if we fix the time-ordering of the surface $\Sigma_{t=0}$ and $\Sigma_{t=1}$ as in Fig.\ref{fig:history}. However, we take the integration range as $-\infty<\text{T}<\infty$  to obtain the well-known Wheeler-Dewitt equation in the path integral formalism. This procedure corresponds to summing over the ordering of the two surfaces too.},
\begin{eqnarray}
&&\int_{-\infty}^{\infty} d \text{T} \underset{z(0)=\epsilon,\ z(1)=z}{\int} [dp_z][dz]\exp\biggl(i \int_{t=0}^{t=1} dt\  (p_z \dot{z} -\text{T} \mathcal{H}_{\Lam})\biggr)\nonumber\\
&=& \mathcal{C}\times \int_{-\infty}^{\infty}  d \text{T}\bra z |e^{-i {\text{T}}\mathcal{H}_\Lam} |\ \epsilon \ket \nonumber\\
&=&\mathcal{C}\times  \bra z | \delta(\mathcal{H}_{\Lam})| \epsilon\ket \nonumber \\
&=&\mathcal{C} \times  \bra z | \delta({\mathcal{H}_\Lam}) \biggl(\int_{-\infty}^{\infty} dE | \phi_{E}\ket \bra \phi_{E} |\biggr) | \epsilon\ket. \label{eq:pathintegral}
\end{eqnarray}
From the first line to the second line, viewing ${\text{T}} \mathcal{H}_\Lam$ as the Hamiltonian, we have used the ordinary relation between the operator formalism and the path integral one, and $\mathcal{C}$ is some constant. In the final line, we have inserted  the complete set \{$|\phi_E\ket$ \}  defined by
\begin{subequations}
\begin{eqnarray}
\bra \phi_E' |\phi_E \ket = \delta(E-E'), \label{eq:normalization}\\
 \mathcal{H}_{\Lam}|\phi_E\ket =E |\phi_E\ket .\label{eq:hamiltonianeq}
\end{eqnarray}
\end{subequations}
Therefore, by using $\phi_{E}(z) \equiv \bra z | \phi_E\ket$, the amplitude can be expressed as
\begin{eqnarray}
\mathcal{C}\times   \phi^*_{E=0}(\epsilon) \phi_{E=0}(z). \label{eq:amplitude}
\end{eqnarray}
In other words, the quantum state of the universe that emerged with size $\epsilon$ is given by
\begin{equation}
\mathcal{C}\times \phi^*_{E=0}(\epsilon)| \phi_{E=0}\ket.\label{eq:universestate}
\end{equation}

We can calculate $\phi_{E}(z)$  in the canonical quantization formalism. 
By replacing $p_z \rightarrow - i \partial / \partial  z$ in the Hamiltonian (\ref{eq:ham}), Eqn.(\ref{eq:hamiltonianeq}) becomes 
\begin{equation}
\sqrt{z}\bigl(\frac{1}{2} \frac{d^2 }{d z^2}-U(z)\bigr) \sqrt{z} \ \phi_E(z)=E \phi_E(z)\label{W-D}.
\end{equation}
Note that for $E=0$ this leads to the  Wheeler-DeWitt equation. However, we need to solve this equation for general $E$ since we should determine the normalization constant of the wavefunction according to \eqref{eq:normalization}.  We rewrite \eqref{W-D} as
\begin{equation}
(-\frac{d^2}{d z^2}-k^2_E(z))\sqrt{z}\phi_E(z) =0,\label{eq:reduced_ODE}
\end{equation}
where 
\begin{eqnarray*}
k_E^2(z) &\equiv& -2U(z)-\frac{2E}{z} \\
&=&9\Lambda
-\frac{9^{1/3}}{z^{2/3}}K_\alpha
+\frac{2M_{matt}}{z}+\frac{2S_{rad}}{z^{4/3}}
-\frac{2E}{z},
\end{eqnarray*}
and apply the WKB method to the function $\sqrt{z}\phi_E(z)$. The solution in
the classically allowed region, $k^2_{E=0}(z)>0$, is  given by a linear combination of 
\begin{eqnarray}
\phi_{E=0}(z) = \frac{1}{\sqrt{\pi}\sqrt{z}\sqrt{k_{E=0}(z)}}\exp(\pm i \int^z dz' k_{E=0}(z') ),\label{eq:general_solution}
\end{eqnarray}
where the normalization is determined by (\ref{eq:normalization}) (see \ref{app:norm}).

We need to specify the boundary condition to determine the solution completely.
As a simple example, if we require $\phi_{E} (0)=0$,\footnote{The boundary condition would be more complicated because the behavior in $z<\epsilon$  is determined by the dynamics near singularity.}  we have
\begin{eqnarray}
\phi_{E=0}(z) = \frac{1}{\sqrt{\pi/2}\sqrt{z}\sqrt{k_{E=0}(z)}}\sin (\int^z dz' k_{E=0}(z') ). \label{eq:positivelambdasolution}
\end{eqnarray}
However, we do not need the details of the solution in the following sections. 

\section{Multiverse Wavefunction and Density Matrix of our Universe}\label{sec:multiverse}

In this section, we construct  the multiverse wave function assuming that all the parent universes have the topology of $S^3$. Here, we mean by the word ``multiverse'' the state with an indefinite number of universes. We then calculate the density matrix of one universe, which is essentially what we observe in our universe.

\subsection{Wave Function of the  Multiverse}

Usually, the  universes which are not connected with ours are irrelevant for us, since they have no effect on our observation. However, when we take the  wormholes into account, these universes interact through them, and all the universes become to have the same coupling constants $\{ \lambda_i\}$, which should be integrated in the path integral.
 
\begin{figure}
\begin{center}
\includegraphics[width=8cm]{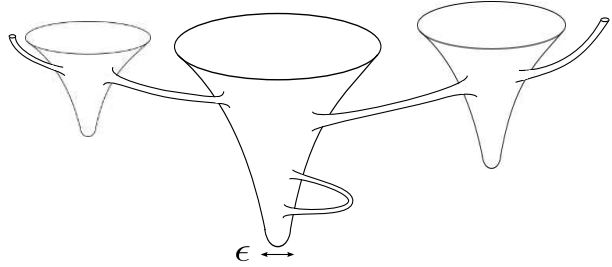}
\caption{A sketch of an example of the  multiverse. Each parent universe emerges with a small size $\epsilon$ by a tunneling process. In this example, the initial state has no baby universes and the final state has two baby universes.}
\label{fig:multiverse}
\end{center}
\end{figure}

\begin{figure}
\begin{center}
\includegraphics[width=7cm]{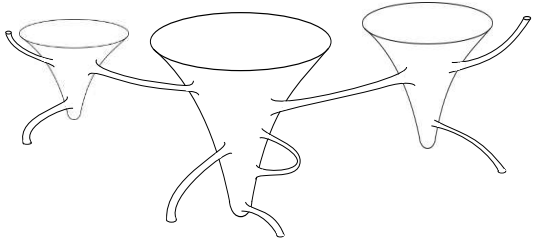}
\caption{A sketch of an example of the  multiverse. In this case, the initial state has some baby universes.}
\label{fig:multiverse_initialbabies}
\end{center}
\end{figure}

In order to construct the quantum state of the multiverse, we need to specify the initial state of the baby universes, which can be expressed as a superposition of the eigenstates of the operators $a_i+a_i^\dagger$, 
\beq
(a_i + a_i^\dagger )|\vec{\lambda} \ket= \lambda_i |\vec{\lambda}\ket,
\eeq
where we have  denoted the set of coupling constants $\{ \lambda_i\}$ by $\vec{\lambda}$.
For example, if there are initially no baby universes  
as in Fig. $\ref{fig:multiverse}$,
the state is given by $  \int  \prod_i d \lambda_i \ e^{- \lambda_i d^{ij} \lambda_j /4} |\vec{\lambda} \ket := |\Omega \ket$, where $|\Omega \ket$ satisfies $a_i |\Omega \ket=0$.\footnote{It might be helpful to regard $a+a^\dagger$ as the position operator $\sqrt{2}x$ of a harmonic oscillator, and recall the ground state of the system $|0\ket$ can be written as $\int dx \ e^{-x^2/2}$ in the $x$-representation. } In general, there may be many baby universes initially (see Figure.\ref{fig:multiverse_initialbabies}),
and the state can be written as $\int \prod_i d \lambda_i \ w(\vec{\lambda}) | \vec{\lambda} \ket$, where $ w$ is a function of $\vec{\lambda}$.

 To write down the multiverse state, we also need the probability amplitude of a universe emerging from nothing, which we denote by $\mu_0$ in analogy of the chemical potential.  Here we assume that all universes are created at the size $\epsilon$.  Together with the factor in (\ref{eq:universestate}), the weight of each universe $\mu$ is given by,
\begin{equation} 
\mu :=\mu_0\times \mathcal{C} \times \phi^*_{E=0}(\epsilon). \label{eq:chemicalpotential}
\end{equation}
A crucial fact is that $\mu$ does not depend on $\Lam$ strongly. This is because  $\phi^*_{E=0}(\epsilon)$ is a smooth function of $\Lambda$ as is seen from (\ref{eq:positivelambdasolution}), and $\mathcal{C}$ arising from the path measure should have nothing to do with $\lambda_i$.

Then, the multiverse wave function can be written as  
\begin{eqnarray}
| \phi_{multi}\ket = \sum_{N=0}^{\infty} | \Phi_N \ket \label{eq:multiversestate}
\end{eqnarray}
where $|\Phi_N \ket$ stands for the $N$-universe state, whose wave function is given by
\begin{eqnarray}
\Phi_N(z_1,\cdots, z_N)= 
\int  d \vec{\lambda} \ \mu^{N} \times \phi_{E=0}(z_1)\phi_{E=0}(z_2)\cdots \phi_{E=0}(z_{N})\ w(\vec{\lambda}) |  \vec{\lambda} \ket \label{eq:multiverse_state_single},
\end{eqnarray}
where 
\beq
  d \vec{\lambda} \equiv \prod_i d \lambda_i.
\eeq

\subsection{Density Matrix of Our Universe} \label{sec:densitymatrix}


We now can obtain the density matrix of our universe by tracing out the other universes and the baby universes, namely $\vec{\lambda}$. Using \eqref{eq:multiverse_state_single}, we can calculate it as
\begin{eqnarray}
\rho(z',z) &=&  \sum_{{N =0}}^\infty  \int \frac{d z_i^{N}}{N ! }
 \ \Phi_{N+1}^* (z', z_1,\cdots,z_N)\Phi_{N+1}(z, z_1,\cdots, z_N)\nonumber \\
&=& \sum_{{N =0}}^\infty  \frac{1}{N ! } \int_{-\infty}^{\infty}  d \vec{\lambda}  \ w(\vec{\lambda})^2  |\mu|^2\phi_{E=0}(z')^* \phi_{E=0}(z)  \times \biggl( \int d z^{''} |\mu \phi_{E=0}(z^{''})|^2 \biggr)^{N}  \nonumber \\
&=& \int_{-\infty}^{\infty}  d \vec{\lambda}  \ w(\vec{\lambda})^2|\mu|^2\ \phi_{E=0}(z')^* \phi_{E=0}(z)\times \exp\biggl(\int dz^{''} |\mu \phi_{E=0}(z^{''})|^2 \biggr),\label{eq:densitymat} 
\end{eqnarray}
where $z\text{\ and}\  z'$ are the size of our universe.
We note that the above integrand  depends on $\{ \lambda_i\}$ through the wave function $ \phi_{E=0}$.


\section{Vanishing Cosmological Constant}\label{sec:vanishing_cc}
In this section, we show that the integrand in \eqref{eq:densitymat} has a strong peak at a point in the $\{ \lambda_i\} $ space where the cosmological constant $\Lambda=\Lambda(\{\lambda_i \})$ becomes very small, which means the cosmological constant problem is automatically solved. We also discuss the possibility of the big fix.

\subsection{Evaluation of the Density Matrix}\label{sec:bf_general}
In this subsection, we examine how the exponent in the density matrix \eqref{eq:densitymat}, 
\beq
\int_0^\infty dz^{''} |\mu \phi_{E=0}(z^{''})|^2, \label{eq:lifetime_closed}
\eeq
depends on $\Lambda$.

First we sketch the potential $U(z)$ in \eqref{eq:ham}.
Again we assume that all the universes have the topology of $S^3$ ($K=1$), so that $U(z)$ is given by
\begin{equation}
2U(z)=-k_{E=0}^2(z)=-9\Lambda
+\frac{9^{1/3}}{z^{2/3}}
-\frac{2M_{matt}}{z}-\frac{2S_{rad}}{z^{4/3}}.\label{eq:closed_pot}
\end{equation}
For large $z$, the leading term is  the cosmological constant $\Lambda$, and the next leading term is the curvature term. We note that only the curvature term is positive, and $U(z)$ has a maximum at one point $z=z_{*}$,
\beq
U'(z_{*})=0.
\eeq

\begin{figure}[htbp]
\begin{center}
\subfigure[$\Lambda<0$]{\includegraphics*[width=.18\linewidth]{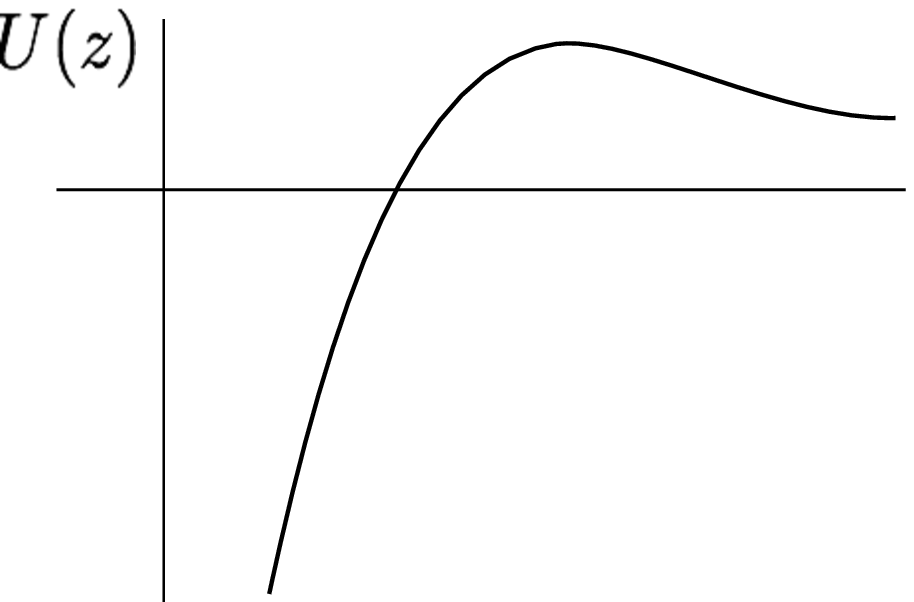}
\label{fig:closed_neg}}
\subfigure[$\Lambda=0$]{\includegraphics*[width=.18\linewidth]
{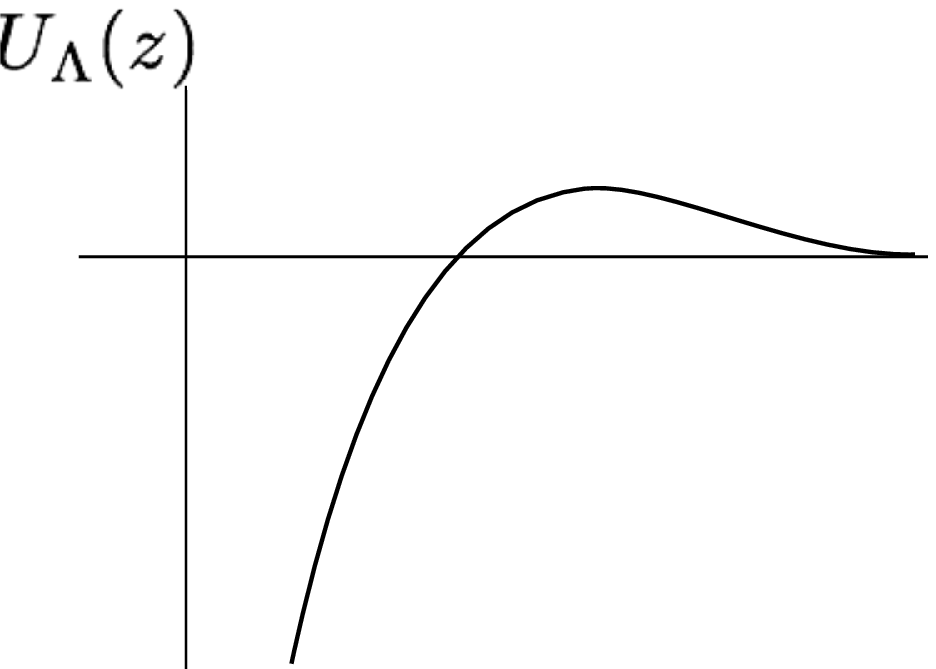}
\label{fig:closed0}}
\subfigure[$0<\Lambda<\Lambda_{cr}$]{\includegraphics*[width=.18\linewidth]{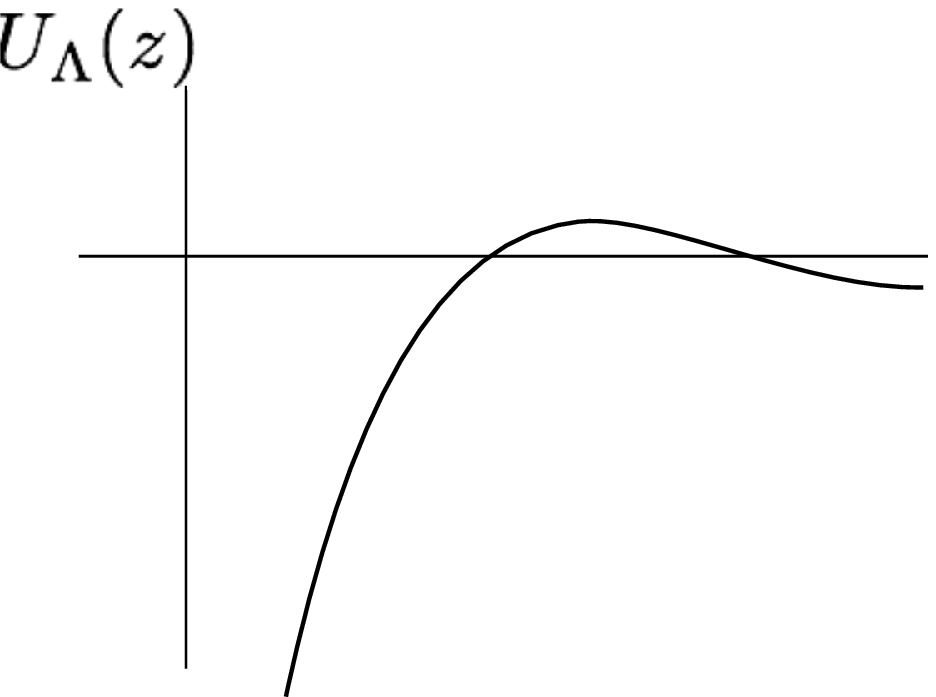}
\label{fig:closed1}}
\subfigure[ $\Lambda=\Lambda_{cr}$]{\includegraphics*[width=.18\linewidth]{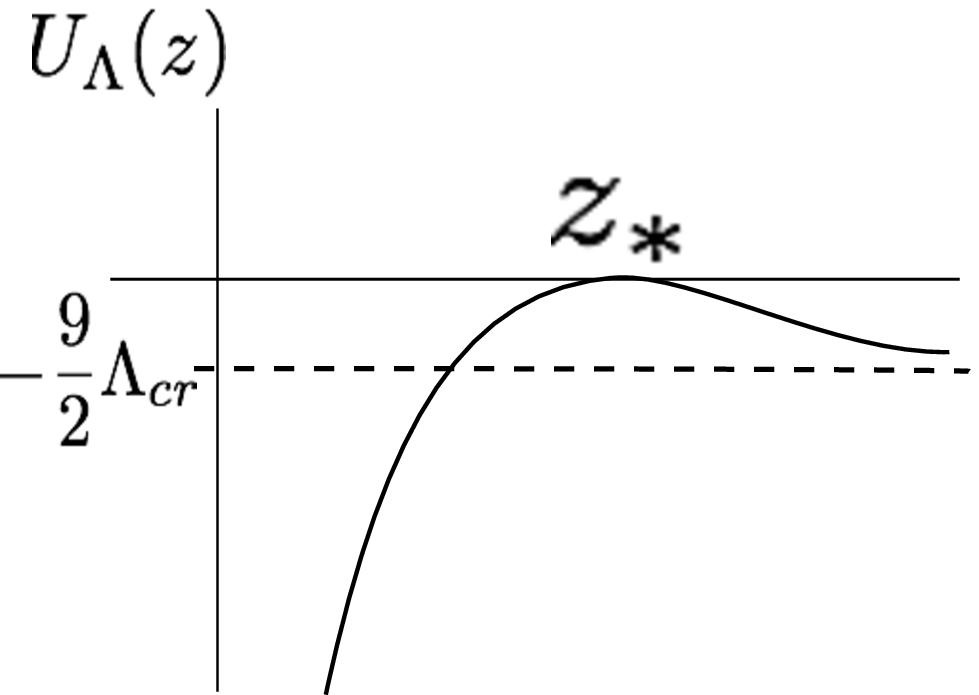}
\label{fig:closed2}}
\subfigure[ $\Lambda>\Lambda_{cr}$]{\includegraphics*[width=.18\linewidth]{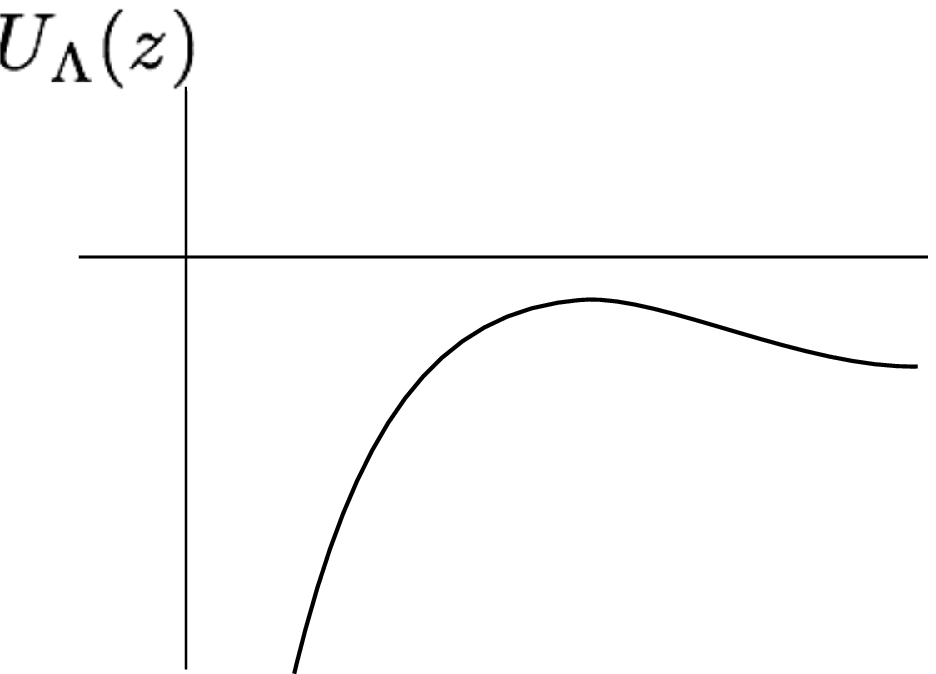}
\label{fig:closed3}}
\caption{As we vary $\Lambda$ from zero to $\Lambda_{cr}\sim \frac{1}{S_{rad}}$, the region where the wave function takes the tunneling suppression becomes shorter. For $\Lambda>\Lambda_{cr}$, there is no suppression.}
\label{fig:closed_series}
\end{center}
\end{figure}

As we vary $\Lambda$  with  $M_{matt}$ and $S_{rad}$ kept fixed,   $U(z)$ changes as in Fig \ref{fig:closed_series}. There is a critical value $\Lambda_{cr}$ at which the maximum becomes zero (see Fig\ref{fig:closed2});  
\beq
U(z_{*}) |_{\Lambda=\Lambda_{cr}}=0.
\eeq
Note that if $\Lambda=\Lambda_{cr}$, three contributions to $U(z)$, the cosmological constant, curvature and energy density coming from matter and radiation, are comparable around $z\sim z_*$. 
The precise values of $z_*$ and $\Lambda_{cr}$ depend on the history of the universe. If all the matter decay into radiation by $z=z_{*}$, we have $M_{matt}=0$, and $\Lambda_{cr}$ is given by \beq
z_{*}= \frac{8S_{rad}^{3/2}}{3},\ \ \  9\Lambda_{cr} =\frac{9^{1/3}}{8 S_{rad}} . {\ \ \rm(for\  radiation\ dominated)} \label{eq:radiation_dominant}
\eeq
On the other hand, if the matter dominates around $z_*$, they are given by 
\beq
z_{*}= \frac{3 M_{matt}}{9^{1/3}},\ \ \ 9\Lambda_{cr} =\frac{1}{3 M^2_{matt}}. {\  \ \rm(for\ matter\ dominated)} \label{eq:matter_dominant}
\eeq

Now we can examine the behavior of $\phi_{E=0}(z)$ in the large-$z$ region, and evaluate the integral \eqref{eq:lifetime_closed}.
If  $\Lambda<0$, the wave function  damps exponentially, and \eqref{eq:lifetime_closed} is finite (see Fig\ref{fig:closed_neg}). On the other hand, if $\Lambda \geq 0$, the wave function does not damp for sufficiently large  $z$, and  \eqref{eq:lifetime_closed} is divergent. Thus, if we introduce a cutoff for large $z$, as we will do below, \eqref{eq:lifetime_closed} takes the maximum for some positive value of $\Lambda$.

Furthermore, if $\Lambda \geq \Lambda_{cr}$, all the region of $z$ is classically allowed, and we can reliably use the WKB solution
\beq
\phi_{E=0}(z)\sim \frac{1}{\sqrt{z k_{E=0}(z)}} \label{eq:wkb_closed},
\eeq
which becomes larger as the momentum $k_{E=0}=\sqrt{-2U}$ becomes smaller. Thus, for $\Lambda \geq \Lambda_{cr}$, the wave function becomes the largest when $\Lambda=\Lambda_{cr}$.
On the other hand, if $0<\Lambda<\Lambda_{cr}$, there is a forbidden region, which suppresses the wave function. The suppression is stronger for smaller $\Lambda$ because the forbidden region becomes larger as we decrease $\Lambda$. 
Thus, we find that \eqref{eq:lifetime_closed} takes its maximum value at
 \beq
 \Lambda=\Lambda_{cr}\label{eq:maximum_rad}.
 \eeq

Next we discuss how the maximum value of \eqref{eq:lifetime_closed} is determined by the amount of radiation $S_{rad}$ or matter $M_{matt}$. If we set $\Lambda=\Lambda_{cr}$, using \eqref{eq:wkb_closed} we have
 \beq
 \int_0^\infty dz^{''} |\mu \phi_{E=0}(z^{''})|^2 \sim  |\mu|^2 \int_0^\infty dz \frac{1}{ z \sqrt{ \Lambda_{cr}}}.
 \eeq
Since this is divergent, we introduce an infrared cutoff $z_{IR}$ and replace $z=\infty$ with $z=z_{IR}$. Then the above integral becomes
 \beq
\int_0^\infty dz \frac{1}{ z \sqrt{\Lambda_{cr}}} \sim \frac{1}{  \sqrt{\Lambda_{cr}}} \log z_{IR},\label{eq:maximum_radiation}
 \eeq
The cutoff $z_{IR}$ should be explained from a microscopic theory of gravity such as string theory. For example, in the IIB matrix model space-times  emerge dynamically from the matrix degrees of freedom, and an infrared cutoff appears effectively, which is proportional to some power of the matrix size\cite{ishibashi1997large,steinacker2007emergent,Kim:2011cr}.

If we consider the case of \eqref{eq:radiation_dominant}, where the curvature term balances with the radiation, \eqref{eq:maximum_radiation} is proportional to $\sqrt{S_{rad}}\log z_{IR}$, and the integrand of the density matrix \eqref{eq:densitymat} behaves as
 \beq
 \exp \biggl( const.  \times \sqrt{S_{rad}} \log z_{IR}  \biggr), \label{eq:case_rad_max}
 \eeq
which has an infinitely strong peak at a point in the $\{ \lambda_i \}$ space where $S_{rad}$ becomes maximum. Here, we have assumed that $|\mu|^2$ does not have a strong dependence on $\{ \lambda_i \}$ because it is determined by the microscopic dynamics of smaller scales than the wormholes.
Thus we have seen that all the couplings $\{ \lambda_i \}$ are fixed in such a way that $S_{rad}$ is maximized. We call it the \textit{big fix} following Coleman. In the original Coleman's argument the enhancement comes from the action itself, or equivalently, the exponential factor in the wave function \eqref{eq:general_solution}, while it comes from the prefactor in our case. We will discuss this meaning in the next subsection. We also note that the big fix applies only to the couplings that are induced by the wormholes.
In particular, the cosmological constant is given by 
\beq
\Lambda=  1/ \underset{ \vec{{\lambda}}  }{\text{max}}\ S_{rad}(\vec{\lambda} ),
\eeq
which is very closed to zero.\footnote{$S_{rad}$ means the amount of the radiation in the whole $S^3$-universe, rather than that in the portion we can observe. Thus, if  the whole universe is large enough, $S_{rad}$ is extremely large. }  We note that $\Lambda$ and $S_{rad }$ appearing above should be regarded as their values at $z=z_*$.

In the other case \eqref{eq:matter_dominant}, where the curvature term balances with the matter, we have $\Lambda_{cr}\sim M_{matt}^{-2}$, and instead of \eqref{eq:case_rad_max} we have  
 \beq
 \exp \biggl( const. \times  M_{matt} \log z_{IR}  \biggr). \label{eq:case_matt_max}
 \eeq
This time, the coupling constants  $\{ \lambda_i \}$ are fixed such that $M_{matt}$ at $z=z_*$ is maximized, and the cosmological constant is given by 
 \beq
\Lambda=  1/ \underset{ \vec{{\lambda}}  }{\text{max}}\ M^2_{matt}(\vec{\lambda} ).
\eeq

In the above mechanism, the curvature term becomes comparable to the cosmological constant around $z = z_*$. On the other hand, observational cosmology tells that the former is much smaller than the latter already in the present universe. Therefore, in order for the scenario to work, the cosmological constant needs to decrease as a function of time by some mechanism such as quintessence models. Then the above argument claims that its asymptotic value is very small.




\subsection{Interpretation of Enhancement at $\Lambda=\Lambda_{cr}$}
In this subsection, we provide an intuitive explanation of the enhancement at $\Lambda=\Lambda_{cr}$ in \eqref{eq:densitymat}. We also argue  that our mechanism works beyond the minisuperspace and the WKB approximation.

First, we recall that  the enhancement of the density matrix comes from the exponent in \eqref{eq:densitymat},
\beq
\int {dz}\ |\phi_{E=0} (z)|^2 \label{eq:time_IR},
\eeq 
and  we have evaluated it by using the WKB solution
\beq
\phi_{E=0}(z)\sim \frac{1}{\sqrt{z k_{E=0}(z)}}.
\eeq
Classically $k_{E=0}(z)$ is the conjugate momentum of $z$, 
\beq
k_{E=0}(z) \sim  \dot{z}/z.
\eeq
Thus, \eqref{eq:time_IR} can be written as
\beq
\int {dz}\ |\phi_{E=0} (z)|^2 =  \int^{z_{IR}}_{\epsilon}{dz}\ \frac{1}{z k_{E=0}(z)}= \int^{z_{IR}}_{\epsilon} \frac{dz}{\dot{z}} \label{eq:lifetime_wkb},
\eeq 
which is nothing but the time it takes for the universe to grow from the size $\epsilon$ to $z_{IR}$. Since we have imposed the cutoff $z_{IR}$ on the size of the universe, a universe with the size larger than $z_{IR}$ does not exist\footnote{Although we have not specified the infrared cutoff precisely, we can simply imagine that when a universe reaches the size $z_{IR}$, it ceases to exist , or it bounces back and starts shrinking towards the size $\epsilon$. }. Thus, \eqref{eq:lifetime_wkb} can be interpreted as the time duration in which the universe exists. We call it the $lifetime$ of the universe, for simplicity.

In fact, we can verify  this interpretation without relying on the WKB approximation. We recall the normalization of the wave function
\beq
\bra \phi_E' |\phi_E \ket = \delta(E-E') \label{eq:normalization_lifetime},
\eeq
which leads to 
\beq
\int dz \ |\phi_{E=0}(z)|^2 \sim \delta(0). \label{eq:normalization_delta}
\eeq
As is usually done in the derivation of Fermi's golden rule, $\delta(0)$ is regarded as the total interval of time, which in our case is naturally interpreted as the duration of the universe.

\begin{figure}[htbp]
\begin{center}
\subfigure[ $\Lambda < \Lambda_{cr}$]{\includegraphics*[width=.25\linewidth]{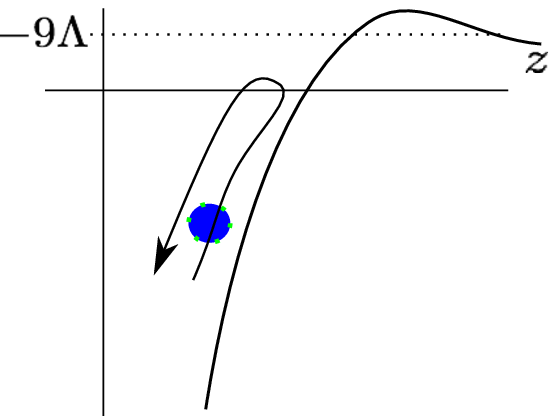}
\label{fig:motion_small}}
\subfigure[$\Lambda\simeq \Lambda_{cr}$]{\includegraphics*[width=.25\linewidth]{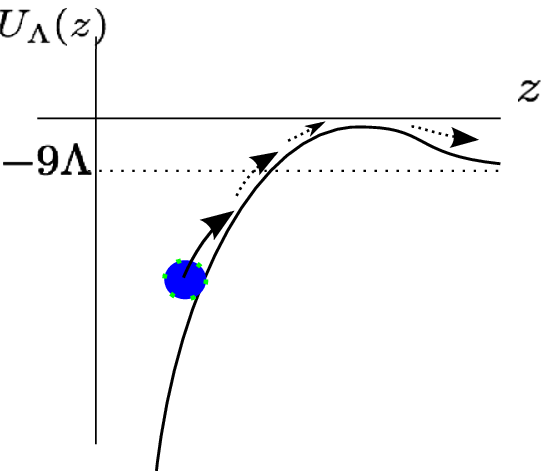}
\label{fig:motion_critical}}
\subfigure[ $\Lambda > \Lambda_{cr}$]{\includegraphics*[width=.25\linewidth]{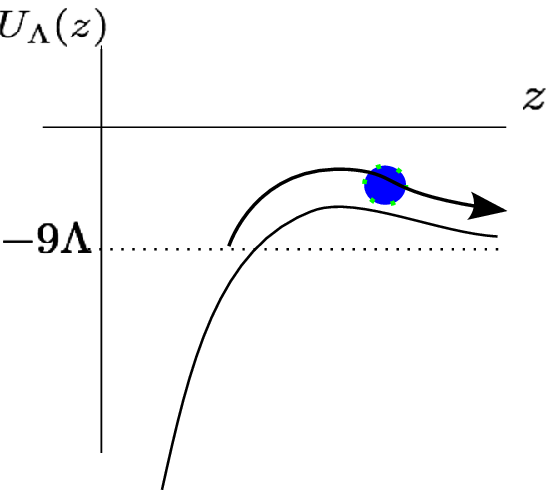}
\label{fig:motion_large}}
\caption{The classical motion is shown for each $\Lambda$. For $\Lambda>\lambda_{cr}$, the universe expands to $z_{IR}$  rapidly, and the lifetime is short. For $\Lambda \simeq \Lambda_{cr}$, it takes long time to reach $z_{IR}$, that is, the lifetime is long.}
\label{fig:classical_motion}
\end{center}
\end{figure}

\begin{figure}[htbp]
\begin{center}
\subfigure[$\Lambda < \Lambda_{cr}$]{\includegraphics*[width=.17\linewidth]{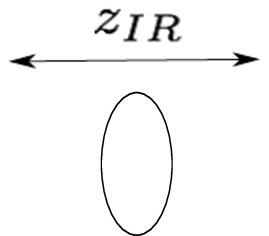}
\label{fig:negative}}
\subfigure[$\Lambda \simeq \Lambda_{cr}$]{\includegraphics*[width=.17\linewidth]{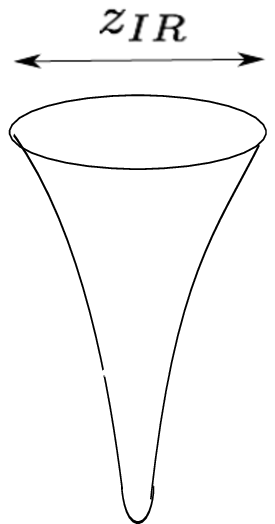}
\label{fig:small_positive}}
\subfigure[$\Lambda>\Lambda_{cr}$]{\includegraphics*[width=.17\linewidth]{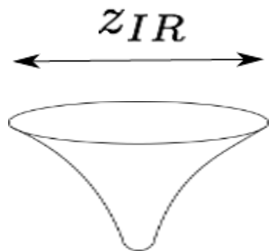}
\label{fig:large_positive}}
\caption{For $\Lambda>\Lambda_{cr}$, as $\Lambda$ varies to $\Lambda_{cr}$, the universe takes more time to expand to the size $z_{IR}$, and the ``lifetime'' becomes longer. For $\Lambda<\Lambda_{cr}$, the universe bounces back to the size zero before reaching $z_{IR}$. }
\label{fig:classical_shape}
\end{center}
\end{figure}

Therefore, what the big fix does is to make the lifetime of the universe as long as possible. Based on this interpretation, we can reproduce the results obtained in the last subsection.
First we note that, for $\Lambda < \Lambda_{cr}$, the universe cannot reach to $z_{IR}$ because of the potential barrier (see Fig.\ref{fig:motion_small}), and collapses back to the size $\epsilon$ and then disappears in finite time  (see Fig.\ref{fig:classical_shape}(a)).\footnote{Quantum mechanically, the universe can reach to $z_{IR}$ after tunneling for $ 0<\Lambda < \Lambda_{cr}$, but because of the tunneling suppression such $\Lambda$ does not contribute much, as we have discussed in the last subsection.}  So we concentrate on the case $\Lambda \geq \Lambda_{cr}$.  As we vary $\Lambda$, the depth of the potential changes as in Figure \ref{fig:classical_motion}. The shallower potential gives the longer lifetime, and thus the lifetime becomes maximum at $\Lambda=\Lambda_{cr}$  (see Fig.\ref{fig:classical_shape}(b) and (c)).

Before closing this subsection, we emphasize the general validity of our mechanism. So far, we have used the mini-superspace approximation, in which only the size of the universe is considered, and the other degrees of freedom such as various fields and inhomogeneous fluctuations of the metric are ignored. 
If we take those degrees of freedom into account, the quantum state of the universe is described not only by $z$, but also by the other degrees $q_i$, and \eqref{eq:time_IR} is replaced by
\beq
\int {dz}\prod_i dq_i \ |\phi_{E=0} (z;q_i)|^2. \label{eq:time_IR_general}
\eeq 
However, if the quantum state of $q_i$'s is normalized to $1$, the integration over $q_i$'s leaves the same integral as the mini-superspace, and again we have $\delta(0)$. 
Therefore, we can say quite generally that the exponent of the density matrix is  the lifetime of the universe. Furthermore, the integral \eqref{eq:time_IR_general} is controlled by large values of $z$, where the evolution of the universe is completely classical. In such late time, the effect of the other degrees of freedom such as gravitons, photons, and protons is simply represented by the energy density in the potential \eqref{eq:closed_pot}, which justifies the analysis we have employed above.

\subsection{Comparison with Euclidean and Other Lorentzian Approaches}\label{sec:comparison}
In this subsection we discuss the difficulty of the Euclidean gravity, and explain how our mechanism is different from the original Coleman's or the subsequent Lorentzian approaches.

\subsubsection{Wrong Sign Hamiltonian}
In order to clarify the problem, we start with a Hamiltonian
\beq
H_{-}= - \frac{p^2}{2}-V(q),\label{eq:wrong}
\eeq
which is the minus of the normal Hamiltonian
\beq
H_{+}=\frac{p^2}{2}+V(q),\label{eq:correct}
\eeq
where $p$ is the canonical momentum of $q$ and   $V(q)$ is a potential. 
Since the Schr\"{o}dinger equation\beq
i \frac{\partial}{\partial t} \Psi(q,t)= H \Psi(q,t)
\eeq
for \eqref{eq:wrong} and \eqref{eq:correct} are simply related by the complex conjugate, they should describe the same physics.
In particular, the tunneling phenomena are the same: When we consider a tunneling process, the wave function should decrease in the direction of the penetration, and the tunneling is exponentially suppressed for both cases.

Next we discuss the Wick rotation of the wrong sign Hamiltonian \eqref{eq:wrong}.
Usually, for the right sign Hamiltonian \eqref{eq:correct}, we rotate the time axis as $t= -i \tau_E$ so that the transition amplitude
\beq \bra q' | e^{-i H_+  t }| q \ket = \bra q' | e^{- H_+ \tau_E }| q \ket =\bra q' | e^{- \tau_E(\frac{p^2}{2} +V(q)) }| q \ket  \label{eq:correctTrans}
\eeq
is well defined.
Note that the rotation in the opposite direction $t= i \tau_E$ does not work because of the bad large-momentum behavior.
On the other hand, for the wrong sign Hamiltonian $H_-$, we should take $t=i\tau_E$ 
\beq \bra q' | e^{-i H_-  t }| q \ket = \bra q' | e^{H_- \tau_E }| q \ket =\bra q' | e^{- \tau_E(\frac{p^2}{2} +V(q)) }| q \ket, \label{eq:wrongTrans}
\eeq
and $t= -i \tau_E$ does not work.

Obviously, \eqref{eq:correctTrans} and \eqref{eq:wrongTrans} are the same, and thus the equivalence of the two systems can be seen also in the Euclidean framework. However, the Wick rotation should be done in such a direction that the transition amplitude is well defined.
In other words, if one applied the naive Wick rotation $t=-i\tau_E$ to the wrong sign Hamiltonian $H_-$, one would have physically unreasonable results.

As an example, we consider the Hamiltonian $H_-$ with $V(q)= \lambda(q^2-q_0^2)^2$ and the transition amplitude 
 \beq 
  \bra q'=+q_0 | e^{-i H_-  t }| q=-q_0 \ket .
 \eeq
If we perform the correct Wick rotation $t=i\tau_E$, the amplitude is given by the ordinary Euclidean path integral as is seen from \eqref{eq:wrongTrans}:
 \beq 
 \bra q'=+q_0 | e^{ H_-  \tau_E }| q=-q_0 \ket = \int \mathcal{D}q \exp \biggl(-\int d\tau \left( \ \frac{1}{2} ( \partial_\tau q)^2+V \right) \biggr).
  \eeq
The one-instanton solution $q_{cl}$ connecting $q=-q_0$ to $q=+q_0$ contributes as
 \beq \bra q'=+q_0 | e^{- H_-  \tau_E }| q=-q_0 \ket\sim Ce^{-S_E[q_{cl}]}+\cdots ,\eeq
where  $S_E[q_{cl}]$ is given by $S_E= \int_{-q_0}^{+q_0} dq \sqrt{2 V(q)}=\frac{4 \sqrt{2}}{3}q_0^3 \sqrt{\lambda}$. This is consistent with the suppression of the tunneling.
On the other hand, if we perform the wrong Wick rotation $t=- i\tau_E$, the amplitude is formally given by an Euclidean path integral for unbounded action
 \beq \bra q'=+q_0 | e^{ - H_-  \tau_E }| q=-q_0 \ket = \int \mathcal{D}q \exp \biggl(\int d\tau \left(  \ \frac{1}{2} (\partial_\tau q)^2+V \right) \biggr).
  \eeq
  Although this path integral is ill-defined, if we naively evaluate it by using the instanton solution $q_{cl}$, we have a wrong answer
\beq \bra q'=+q_0 | e^{- H_-  \tau_E }| q=-q_0 \ket\sim Ce^{S_E[q_{cl}]}+\cdots .
\eeq
This would indicate that the tunneling is not suppressed but enhanced exponentially. However, as we have discussed above, we do not regard it as true.

\subsubsection{Case of the Quantum Gravity}
 We now turn to the case of quantum gravity, whose Hamiltonian is schematically given by 
\beqa
H= \frac{1}{2a}[- \Pi_a^2 + f(a)\Pi_{trans}^2]+ \cdots,
\eeqa
where $\Pi_{trans}$ stands for the canonical momentum of transverse modes of the metric, and $f(a)$ is a positive function of $a$. We note that the signs in front of  $\Pi_a$ and $\Pi_{trans}$ are opposite. 
The dots represent various matter and gauge fields, which have the same sign as the transverse modes.
Thus, if we perform the standard Wick rotation $t= -i \tau$ to make the transverse and matter sectors well-defined, we lose control of the fluctuation of the conformal mode. On the other hand, if we take $t=+i\tau$ to avoid it, then the transverse and matter sectors become divergent. Thus, the time axis cannot be rotated in any direction, and the Euclidean gravity obtained by a simple Wick rotation is problematic.\footnote{There is some argument that the analytic continuation of the conformal mode might cure the problem\cite{gibbons1978path}. Here we do not consider this possibility since the physical meaning of the complexified scale factor is not clear.} 

In order to clarify the origin of the confusions about the Euclidean gravity,
we consider the tunneling nucleation of the initial universe. 
The Hamiltonian  in the mini-superspace is given by
\beq
H_{grav}= \frac{1}{2a}(- \Pi_a^2 -a^2 + \rho_{vac} a^4),
\eeq
where $\rho_{vac}$ is the vacuum energy of the universe in the planck epoch.
 Classically, the evolution of $a(t)$ is given by solving $H_{grav}=0$, and in  quantum mechanics, it is  promoted to the constraint on the wavefunction of the universe,
\beq
(\frac{\partial^2}{\partial a^2} -a^2 + \rho_{vac}a^4) \Psi(a)=0.
\eeq
As Vilenkin showed by using the WKB analysis\cite{vilenkin1984quantum}, the tunneling probability $\mathcal{P}$ from $a=0$ to $a= 1/\sqrt{\rho_{vac}}$ is given by
\beq
\mathcal{P}_{WKB} \propto e^{-\frac{2}{3 \rho_{vac}}}.\label{eq:Vilenkin}
\eeq
This result can be obtained  in the Euclidean formalism, if we apply  the  Wick rotation correctly, $t= i\tau_E$, as we have discussed in the previous subsection. Then, the bounce solution $\bar{a}(\tau_E)$ is given by
\beq
\bar{a}(\tau_E)= \frac{1}{\sqrt{\rho_{vac}}} \cos (\sqrt{\rho_{vac}}\tau_E),
\eeq
and, for this solution, the Wick rotated action is evaluated as
\beq
S_E[\bar{a}] = \int d\tau_E  \frac{1}{2}\bar{a}\biggl(1+ (\frac{\partial \bar{a}}{\partial \tau_E})^2 -\rho_{vac} \bar{a}^2\biggr)= \frac{2}{3\rho_{vac}},
\eeq
from which we obtain  the tunneling probability $\mathcal{P}$ as
\beq
\mathcal{P}\propto \exp(-S_E)= \exp(-\frac{2}{3\rho_{vac}}).\label{eq:tunnel}
\eeq
We can thus recover  \eqref{eq:Vilenkin}, and there is no enhancement as $\rho_{vac}\rightarrow +0$.

On the other hand, if we performed the Wick rotation in the wrong direction 
$t=-i\tau_E$, which is the case of the ordinary Euclidean gravity,  we would obtain an enhancement instead of the suppression,
\beq
\mathcal{P}= \exp(S_E)=\exp(\frac{2}{3\rho_{vac}}),
\eeq
which states that the bigger universe is more likely produced via the tunneling.
It seems that this picture is accepted in the original Coleman's and some of the subsequent works, and used to discuss the possibility of the double exponential form  $\exp(\exp(\frac{2}{3\rho_{vac}}))$ in the multiverse. However, as we have discussed, we do not accept this picture, and we simply trust the results of the Lorentzian gravity, in which the tunneling is suppressed. 
Therefore, we do not claim the double exponential form, and instead we have shown a different origin of the enhancement, which leads to  \eqref{eq:case_rad_max} or \eqref{eq:case_matt_max}.

\subsubsection{Enhancement in the Lorentzian gravity}

Here we discuss how our enhancement mechanism is related to the probabilistic interpretation of the Wheeler-DeWitt (WDW) wave function,
 and compare our mechanism with the other authors'.

First, we emphasize that our enhancement mechanism has a completely different origin from Coleman's original idea; he obtained the enhancement at $\Lambda=0$ from the path integral itself, which is evaluated by the 4-sphere solution as
\beq
 \int   \mathcal{D}g\ e^{-S_E}\sim   e^{\frac{1}{\Lambda} }.
\eeq
We think this is fake as we have discussed in the previous subsection.
On the other hand, our enhancement mechanism has nothing to do with the value of the path integral. In fact, by using  \eqref{eq:pathintegral}, the amplitude of a universe emerging with $z_i=\epsilon$ and terminating with  $z_f=\epsilon$ is evaluated as
\beq
\int dT \langle \epsilon |e^{- iHT}  | \epsilon\rangle \sim  |\mu|^2 \phi_{E=0}( \epsilon) \phi^*_{E=0}( \epsilon),
\eeq
which is not particularly enhanced. 
 
 Even though the path integral itself does not have enhancement,  it arises from the probability measure of the WDW wavefunction. In this paper, we have simply assumed that the absolute value squared of the wave function gives the probability density\cite{hawking1986operator,moss1984wave}\footnote{For a review of the various interpretations, see for example \cite{unruh1989time}.}. More specifically, the multiverse state \eqref{eq:multiversestate} is the superposition of  $N$-verse states each of which consists of $N$ universes with sizes $z_1,\cdots,z_N$, and coupling constants $\{\bf{\lambda}\}$, and we interpret 
\beq
|\Phi(z_1,\cdots,z_N)|^2 d{\bf{\lambda}}  dz_1 \cdots dz_N =|\mu|^{2N} \prod_{i=1}^N |\phi_{E=0}(z_i)|^2 dz_i d{\bf \lambda} \label{eq:Nverse}
\eeq
as the probability of finding $N$ universes with the sizes $z_i\sim z_i +dz_i$($i=1,\cdots, N$) and the coupling constants $\{{\lambda}\}$.\footnote{Here, we omit the weight of the coupling constants, $w(\lambda)$ in \eqref{eq:densitymat} since it does not play any important role in the argument.  }

Although this probability measure is a straightforward generalization of the ordinary quantum mechanics,
\beq
|\phi(x,t)|^2 dx,
\eeq
there is some criticism. If we evaluate the normalization integral 
\beq
\int dz |\phi_{E=0}(z)|^2 \label{eq:div},
\eeq  
we find a divergence for large $z$. It essentially comes from the integral over time $T$ in the path integral \eqref{eq:pathintegral}, which makes the universe a superposition of $z$. Thus, the measure \eqref{eq:Nverse} appears to correspond to the following probability  measure in the ordinary quantum mechanics
\beq
 |\phi(x,t)|^2 dx dt,
\eeq
whose integral is obviously divergent since $\int |\phi(x,t)|^2 dx$ is constant in time.

However, we adopt the probability measure \eqref{eq:Nverse} as the most natural one. The reason is the following: Suppose  we perform a numerical simulation of some microscopic model that realizes  the emergence of the multiverse. Then, every time we make an observation, we find an $N$-verse which consists of $N$ universes with various sizes. Therefore, we are naturally lead to consider the ensemble  of $N$ universes with the probability \eqref{eq:Nverse}. The divergence of  \eqref{eq:div} practically does not cause any problems in the process of the simulation. As we have mentioned, the infrared cutoff $z_{IR}$ is naturally introduced, for example, as the size of the matrix when we design the spacetime geometry by matrixes, or the number of the simplexes in the dynamical triangulation.

In order to understand how the enhancement arises from the measure \eqref{eq:Nverse}, we first consider the single universe state. The WDW wave function of the universe $\mu \phi_{E=0}(z)$ represents the superposition of various universes with size $z$. As we have seen around \eqref{eq:normalization_delta}, the measure $|\mu \phi_{E=0}(z)|^2 dz$ can be interpreted as the probability distribution of the time $T$
 that has passed after the universe emerged,
 \beq
 |\mu \phi_{E=0}(z)|^2 dz \sim |\mu|^2 dT.
 \eeq
If we integrate it over $z$, we find that each universe has the weight 
\beq
\int dz |\mu \phi_{E=0}(z)|^2 \sim |\mu|^2 T_{\lambda}.\label{eq:T}
\eeq
Here, $T_{\lambda}$ is the lifetime of the universe, which depends on the coupling constants $\{ \lambda \}$.
Thus, our probability measure counts the universe with the weight $|\mu|^2 T_{\lambda}$.  
Similarly,  the $N$-verse state is the superposition of the states each of which consists of $N$ universes which were created at various times.
Therefore, \eqref{eq:Nverse} is equal to
 \beq
|\mu|^{2N} d T_1 \cdots dT_N, 
\eeq
where we consider that the i-th universe was created time $T_i$ before the observation.
 If we integrate \eqref{eq:Nverse} over the sizes, the $N$-verse state is counted with the weight 
\beq
\frac{1}{N!}|\mu|^2 T_{\lambda},
\eeq
where $N!$ is the symmetry factor. When we evaluate the density matrix \eqref{eq:densitymat}, the lifetime $T_{\lambda}$ becomes exponentiated to
\beq
\exp\left(|\mu|^2 T_{\lambda} \right)  \tag{\ref{eq:time_IR}}
\eeq
after summing over the number of the universes. Thus, our enhancement mechanism essentially 
comes from the probability measure, which counts each universe  with the weight of the lifetime.


We expect the big fix occurs in such a way that the lifetime is maximized.
This point is completely different from the earlier works based on Lorentzian gravity \cite{rubakov1988third, strominger1989lorentzian,fischler1989quantum,cline1989does}, 
 In particular, our mechanism has nothing to do with the initial tunneling amplitude $\mu$. As we have seen from \eqref{eq:tunnel}, $\mu$ in general depends on the various coupling constants $\{\lambda \}$  at the planck scale. However, what determines the lifetime of the universe  is not  the microscopic parameters themselves but the parameters at the low energy scale, such as the renormalized cosmological constant and the Higgs mass, and so on, and there is no reason that $\mu$ has a strong dependence on such low energy quantities. Thus, the tunneling amplitude $\mu$ does not play an important role in the big fix.

\section{The Big Fix and the Gauge Hierarchy Problem}\label{sec:bigfix}
One of the notorious problems of the standard model is the gauge hierarchy problem, which arises from the quadratic divergence of  the Higgs  mass. In this section, assuming that the wormhole effect induces the parameters of the Higgs potential, the VEV  $v_h$ and the quartic coupling $\lambda_h$, we examine the possibility that the hierarchy problem is solved by the big fix. Here we take, as the low energy effective theory, the ordinary standard model with the proton decay at the GUT scale, and fix the gauge and the yukawa couplings to the observed values.  In order to discuss the big fix, we need to know  the universe in the future. Here we assume that the curvature term balances with the radiation after the baryons decay, which corresponds  to  the case of Fig.\ref{fig:closed2} and Eqn.\eqref{eq:case_rad_max}.
Such a universe is realized if, for example, the following conditions are satisfied: 

$Condition\ 1$. The cosmological constant is time-dependent and decreases to the  asymptotic value before the proton decay. 

$Condition \ 2$. The lifetime of the dark matter  is much shorter than  that of protons.  

$Condition\ 3$. The curvature balances with the energy density while the decay products of baryons being relativistic.

\begin{figure}
\begin{center}
\includegraphics[width=10cm]{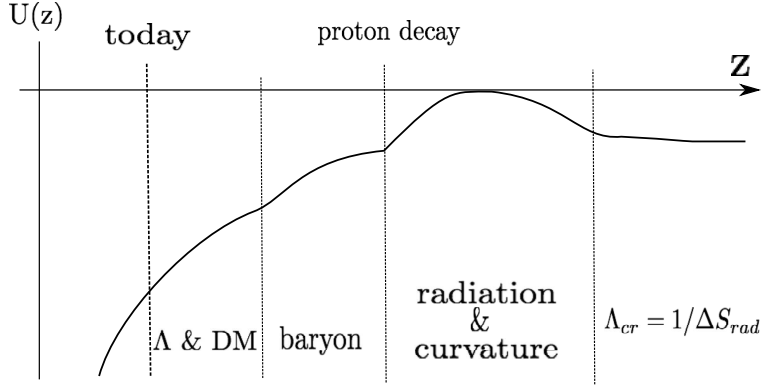}
\caption{A sketch of the potential. 
}
\label{fig:potential_bf}
\end{center}
\end{figure}

A comment is in order on the above conditions. If they are satisfied, the universe evolves like in Fig.\ref{fig:potential_bf}.   $Condition  \ 1$ and $2$ ensure that the cosmological constant and the dark matter  become irrelevant in the energy density, and so the baryons dominate the energy density. However, around the proton lifetime, the baryons decay and  the radiation such as relativistic electrons are produced after the decay. Finally, as the universe expands, the leptons become non-relativistic, namely become matter, due to the red-shift. As we have discussed in Section \ref{sec:bf_general}, we need to specify in which stage the curvature term becomes comparable to the energy density. $Condition \ 3$ claims that it happens in the third stage as is shown in Fig.\ref{fig:potential_bf}. In general, as we will discuss in  \ref{app:e-folds}, the e-foldings of the initial inflation determines when it happens, and the above scenario corresponds to the values given by \eqref{eq:e-folds}.

In Section \ref{sec:bf1}, we discuss how the proton decay determines $S_{rad}$ in the far future, and we write it in terms of  the proton mass $m_p$, the total baryon number $N_b$, and the pion mass $m_\pi$.  In Section \ref{sec:bf2},  we will analyze how these quantities depend on $\lambda_h$ and $v_h$ and at what values they are fixed. In Section \ref{sec:bf3}, we  discuss the strong CP problem.



\subsection{Proton Decay and the Radiation}\label{sec:bf1}

We denote the proton decay rate by $\Gamma_p$, and its inverse by $\tau_p$. When the protons decay into radiation at some large scale factor $a_p \simeq a(\tau_p)$, the continuity of the energy density  leads to the following relation
\beq
\frac{\Delta M_{matt}}{a_p^3}= \frac{\Delta S_{rad}}{a_p^4} \Longrightarrow \Delta S_{rad}= \Delta M_{matt} \times a_p, \label{eq:decay_conservation}
\eeq
where $\Delta M_{matt}$ is the contribution of protons to $M_{matt}$, and $\Delta S_{rad}$ is the radiation amount produced by their decay.
 Because $\Delta M_{matt}$ is expressed  as $N_b \times m_p$, the second equation of \eqref{eq:decay_conservation} becomes\beq
\Delta S_{rad}= m_p \times N_b  \times a_p. \label{eq:delta_rad}
\eeq

We  assume that the cosmological constant $\Lambda$ decreases so rapidly that the universe is mattar-dominated in most of the time until the proton decay.
Then, the Friedman equation $(\dot{a}/a)^2=\frac{M_{matt}}{a^3}$ determines the evolution of the scale factor as
\beq
a_p\propto \Delta M_{matt}^{1/3} \tau_p^{2/3}=(m_p N_b)^{1/3} \tau_p^{2/3},
\eeq
and \eqref{eq:delta_rad} becomes
\beq
(\Delta S_{rad})^{3/2} = N_b^{2} m_p^{2}  \tau_p.\label{eq:rad_pro}
\eeq

$\tau_p$ can be estimated as follows. The effective interaction which induces the proton decay is given by $\bar{e} \pi p$ with the coupling constant $g / M_x^2$, where $M_x$ is the GUT scale and $g$ has the mass dimension two, $g \sim \Lambda_{QCD}^2$.
Using the formula of the decay rate,
\beq
\Gamma_p \sim \frac{1}{2 m_p} \biggl( \int \frac{d^3 p_\pi}{(2\pi)^3}\ \int \frac{d^3 p_e}{(2\pi)^3}\biggr)|\mathcal{M}|^2 (2\pi)^4 \delta^{(4)}(p_p-p_\pi-p_e),
\eeq
where $\mathcal{M}$ is the matrix element of the decay process,
we have 
\beq
\Gamma_p = \tau_p^{-1}\propto g^2 m_p \biggl(1-\frac{m_\pi^2}{m_p^2}\biggr)^2,
\eeq
and \eqref{eq:rad_pro} becomes
\beq
(\Delta S_{rad})^{3/2} \propto N_b^{2} g^{-2} m_p \biggl(1-\frac{m_\pi^2}{m_p^2}\biggr)^{-2}.\label{eq:rad_pro_kansei}
\eeq

\subsection{The Big Fix of the Higgs Parameters}\label{sec:bf2}

\subsubsection{The Higgs Vacuum Expectation Value $v_h$}
 
 Before discussing the big fix of $v_h$, we note that, since we regard the yukawa couplings $y_{u,d}$ as constants  in our argument, we can consider  the current quark mass $m_{u,d}$, instead of $v_h$:
\beq m_{u,d} = v_h y_{u,d}.\eeq
Then, what we want to know is the value of $m_{u,d}$ that maximizes the radiation amount $\Delta S_{rad}$ in \eqref{eq:rad_pro_kansei}.
$N_b$ does not  depend on $v_h$ much if we assume the leptogenesis in which the baryons  are mainly produced in the energy scale  much higher than $v_h$. Therefore, we concentrate on the remaining quantities
\beq
 g, \ m_p, \ m_\pi. \label{eq:couplings}
 \eeq 

If $m_{u,d} >> \Lambda_{QCD}$, a simple quark counting and the dimensional analysis tell us that the masses and the coupling constant $g$ are given by 
\beq
m_{p}\sim 3 \times m_{u,d}, \  m_{\pi}\sim 2 \times m_{u,d}, \ g\propto m_{u,d}^2,
\eeq 
which means that $\Delta S_{rad}$ is a decreasing function for large $m_{u,d}$. We thus examine the possibility that $\Delta S_{rad}$ becomes maximum at some small value of $m_{u,d}$.

We need an expression of  the quantities \eqref{eq:couplings} for  small $m_{u,d}$.  Firstly, the proton and pion masses are given by
\begin{eqnarray}
m_\pi^2&=&\alpha M_p^{(0)} m_{u,d},\nonumber\\
m_p&=& M_p^{(0)}+ 3 \beta m_{u,d}, \label{eq:mass}
\end{eqnarray}
where $M_p^{(0)}$ is the proton mass in the absence of the current quark mass, and $\alpha$ and $\beta$ are some numerical parameters.\footnote{Naively, proton mass is expected to be written as $m_p= M_p^{(0)}+ 3 m_{u,d}$. However, turning on non-zero $m_{u,d}$ affects the chiral condensation. We express the total effect by the parameter $\beta$.} Both of $\alpha$ and $\beta$ are determined by the dynamics of massless QCD, and are independent of $m_{u,d}$. Experimentally we have
\beq
 M_p^{(0)} \approx 910 \text{~MeV}, \ m_{u,d} \approx 5-10 \text{~MeV} ,
\eeq
and so $\alpha$ takes  some value around $2 < \alpha<4$. 

On the other hand, since $g$ has the mass dimension two, it can be expanded in $\frac{m_{u,d}}{M_p^{(0)}}$ as follows:
\beq
g \propto (M_p^{(0)})^2\left( 1+ 3 \beta \kappa \frac{m_{u,d}}{M_p^{(0)}} \right),\label{eq:kappa}
\eeq
where $\kappa$ is some parameter around $0<\kappa<2$.\footnote{This range of $\kappa$ seems reasonable if we rewrite  \eqref{eq:kappa} as 
 \beq
g \sim m_p^\kappa (M_p^{(0)})^{2-\kappa}.
\eeq
} 

Substituting \eqref{eq:mass} and \eqref{eq:kappa} into \eqref{eq:rad_pro_kansei}, we find that 
\beq
(\Delta S_{rad})^{3/2} \propto \bigl(1+ 3\beta  x \bigr)^{-2\kappa +1} \biggl( 1-  \frac{\alpha  x}{(1+3\beta  x)^2}\biggr)^{-2}, \label{eq:before_expand}
\eeq
where we have introduced $x\equiv \frac{m_{u,d}}{M_p^{(0)}}$. \eqref{eq:before_expand} can be expanded as
\begin{eqnarray}
(\Delta S_{rad})^{3/2}& \propto& 1+\biggl(2 \alpha -6 \kappa +3 \beta \biggr)x  +\mathcal{O}(x^2).
\label{eq:expand}
\end{eqnarray}
which indicates $\Delta S_{rad}$ is a increasing function for small $m_{u,d}$ if  
 \beq
2 \alpha -6 \kappa +3 \beta >0.
\eeq
If it is the case, since we have seen $\Delta S_{rad}$ is decreasing for large $m_{u,d}$, we can conclude that  $\Delta S_{rad}$  takes its maximum at some small $m_{u,d}$.

In order to determine the concrete value of $x$, we need the second order term in \eqref{eq:expand}, and more precise analysis of QCD is required. It would be very interesting to see whether or not $\Delta S_{rad}$ really takes its minimum at the experimental value of $x$, $ \frac{5}{910} < x  <\frac{10}{910}$. If it works, the big fix fixes $m_{u,d}$ to $ 5 \sim 10 \rm{MeV}$, which implies the  Higgs VEV to be
 \beq
 v_h \sim \mathcal{O}(\text{100GeV}).
 \eeq

\subsubsection{The quartic coupling constant and the Higgs mass}
Assuming that $v_h$ is correctly fixed at $v_h \sim 246 \text{GeV}$, we next discuss  the quartic coupling constant $\lambda_h$, and predict the Higgs mass. 

The $\lambda_h$-dependence of $\Delta S_{rad}$ is quite simple because $\lambda_h$ enters only $ N_b$ in \eqref{eq:rad_pro_kansei}.\footnote{Although we have neglected $v_h$-dependence of $N_b$ in the discussion of the big fix of $v_h$, we can not ignore $\lambda_h$ in $N_b$ because $\lambda_h$ only appears in $N_b$ in \eqref{eq:rad_pro_kansei}.} 
Since in the leptogenesis scenario most of the baryons are produced swiftly in the symmetric phase,  the baryon number does not depend on the Higgs parameters strongly. However, if we make $\lambda_h$ smaller, the period of the symmetric phase becomes longer. Thus, the number of the baryons $N_b$ becomes slightly increased. Therefore, $N_b$ is  a decreasing function of $\lambda_h$, and 
 smaller $\lambda_h$ dominates in the density matrix \eqref{eq:densitymat}. 

However, it is well known that there is a lower bound for $\lambda_h$ from a stability of the Higgs potential. This bound corresponds to the case that the coupling $\lambda_h$ vanishes at the Planck scale, or wormhole scale.\footnote{We assume that the wormhole size is almost equal to the Planck scale. 
} Thus,  $\lambda_h$ is fixed to this lower bound by the big fix. As shown in \cite{holland2005triviality}, the corresponding Higgs mass $m_h$ is around
  \beq
  m_h \simeq140\pm 20 \text{~GeV}. \label{eq:higgs_pred}
  \eeq

We note that while we need  some assumptions of cosmology in order to discuss $v_h$, the argument of the Higgs mass seems relatively generic.
 \eqref{eq:higgs_pred} can be derived only by assuming that the Higgs VEV is $v_h \simeq 246 \rm{GeV}$ and that the energy density of the universe is a decreasing function of $\lambda_h$. 

\subsection{Strong CP problem}\label{sec:bf3}
So far we have assumed that the CP violating phase $\theta$ is vanishing since there is an experimental upper bound $\theta < 10^{-11}$. We can also discuss the strong CP problem by examining how the non-zero deviation of $\theta$ influences the  radiation amount $\Delta S_{rad}$ in \eqref{eq:rad_pro_kansei}.

Fortunately, we can make an argument without knowing the specific $\theta$-dependence of $\Delta S_{rad}$. The baryon number $N_b$ does not depend on $\theta$ since $N_b$ is determined at much higher energy, and  the remaining quantities, $m_p$, $m_\pi^2$, $g$, should respect a reflection symmetry due to the CP transformation:
  \beq
  \theta \rightarrow -\theta. \label{eq:CP_theta}
  \eeq
Strictly speaking, the real CP transformation flips the sign of the CKM phase as well as $\theta$. However, the reflection of $\theta$ is an almost exact symmetry in the hadronic scale, which is much lower than the weak scale. Thus, $\Delta S_{rad}$ must be an even function of $\theta$, and we have only two possibilities: the point $\theta=0$ maximizes or minimizes $\Delta S_{rad}$ ( at least locally). 
If the former is the case, and $\theta=0$ is the global maximum, $\theta$ is fixed to zero by the big fix. It would be very interesting to examine by QCD whether it is really the case or not.

We note that this argument is highly generic because it relies only on symmetry, and so we can still make a similar argument even when a cosmology other than that we assumed in this section is realized.




\section{Universes with Different Topologies}\label{sec:others}
So far we have only discussed  closed universes with topology $S^3$ ($K=1$). In this section, we study the universe with other topologies. 
 We first discuss the case that all the universes are flat ($K=0$), and compute the density matrix. We find that  it has a strong peak at $\Lambda=0$. We then consider the case that all the universes are open ($K=-1$). 
  
 Finally, we construct the density matrix in the case where various topologies are allowed in the multiverse state. We will find that the flat universes dominate in the density matrix.

 \subsection{Flat Universes}
We consider the case that the multiverse consists of flat universes.  
For $K=0$, the potential $U(z)$ is  given by 
\begin{equation}
2U(z)=-9\Lambda
- \frac{2M_{matt}}{z}-\frac{2S_{rad}}{z^{4/3}}.
\end{equation}

\begin{figure}[htbp]
\begin{center}
\subfigure[$\Lambda>0$]{\includegraphics*[width=.30\linewidth]{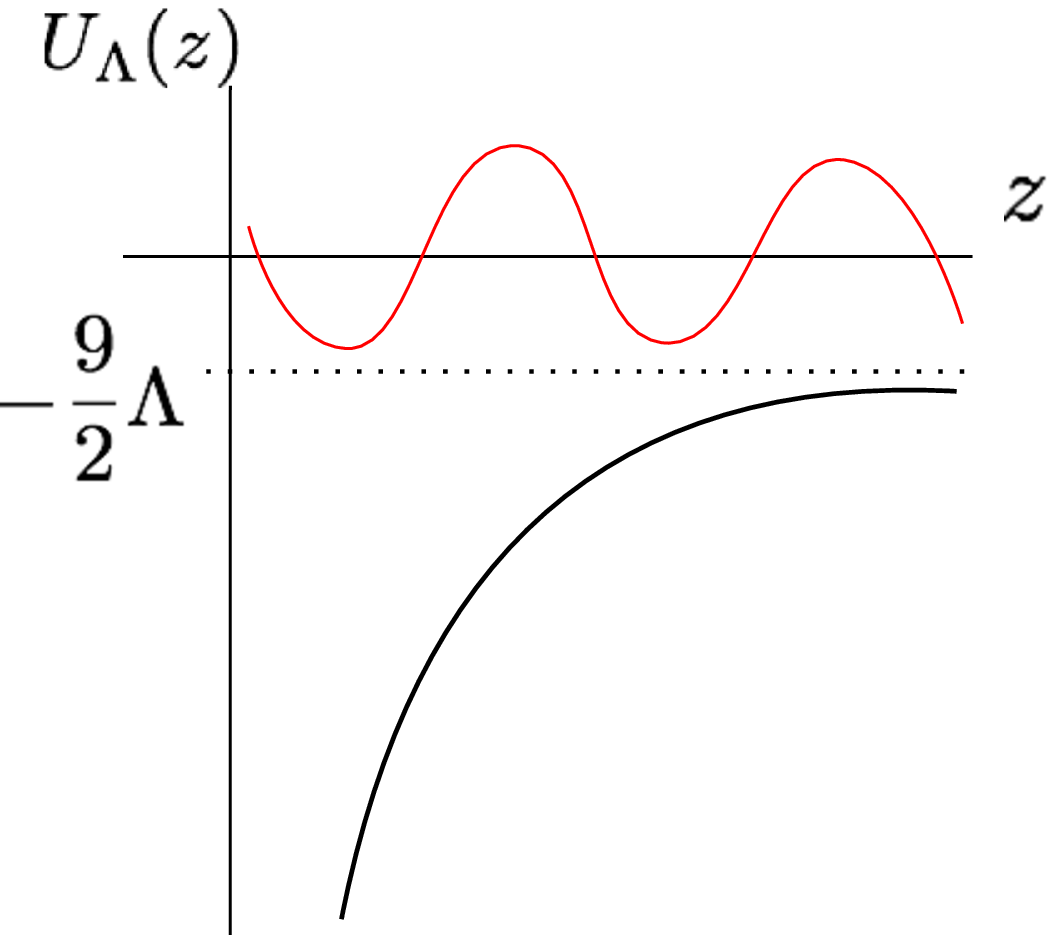}
\label{fig:flat_positive}}
\subfigure[$\Lambda<0$]{\includegraphics*[width=.30\linewidth]{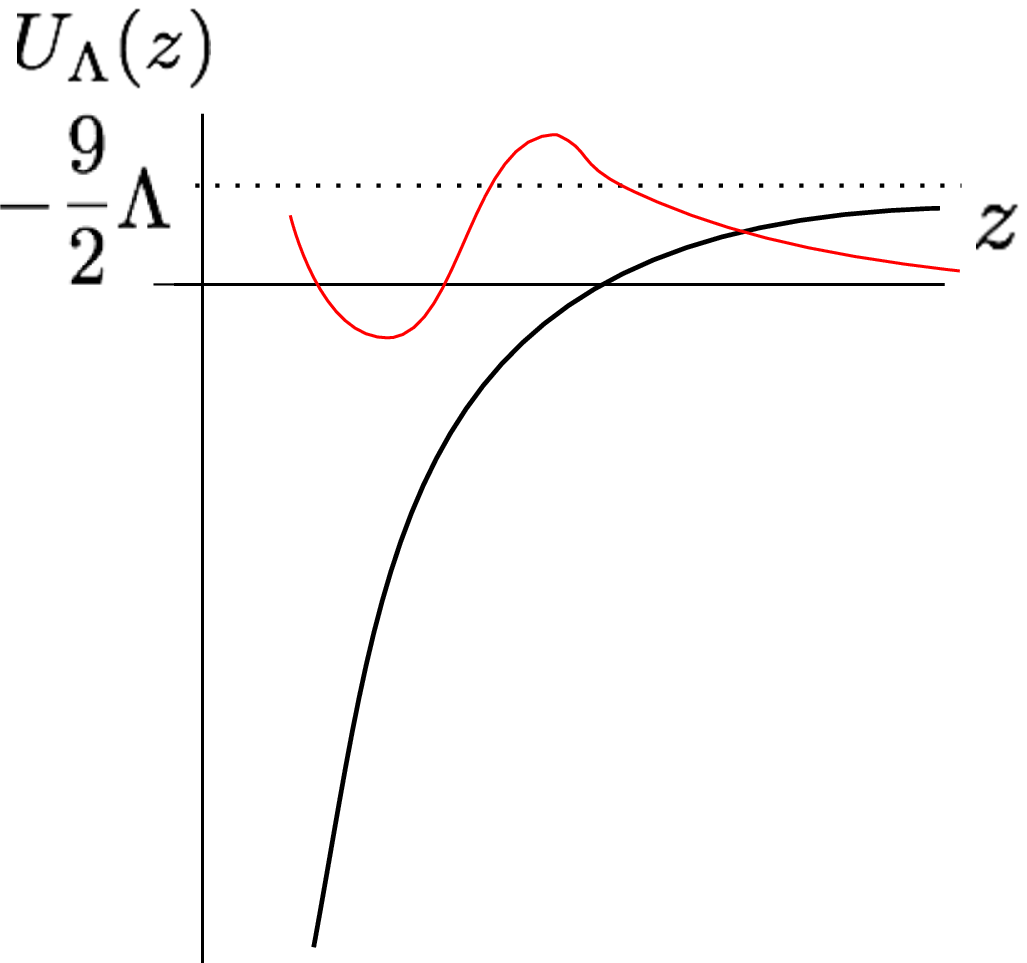}
\label{fig:flat_negative}}
\end{center}
\caption{The potential $U(z)$ for the flat universe. The solid line is the potential, and the dashed line is its asymptotic value $-9\Lambda/2$. The colored line represents a typical form of the wavefunction $\phi_{E=0}(z)$. }
\label{fig:flat}
\end{figure}

If $\Lambda>0$, the whole region of $z$ is classically allowed, and the integral of the wave function can be evaluated as follows by using the solution (\ref{eq:positivelambdasolution}) with $K=0$:
\begin{equation}
\int dz |\phi_{E=0}(z)|^2 = \int dz \frac{2}{\pi z k_{E=0}(z)} \sin^2( \int  k_{E=0}(z')dz' )),
\end{equation}
which is divergent and we regulate it by an infrared cutoff $z_{IR}$  as before.
Then it behaves as 
\begin{equation}
\int^{z_{IR}} dz |\phi_{E=0}(z)|^2 \sim \frac{1}{\sqrt{9\pi^2\Lam}} \log z_{IR} + \cdots.
\end{equation}
On the other hand, if $\Lambda<0$, $\phi_{E=0}(z)$ damps exponentially for large $z$, and the integral gives a finite value. Therefore, this region can be neglected in the density matrix.

Then we obtain the following density matrix \eqref{eq:densitymat},
\begin{eqnarray}
\rho \sim \int_0^\infty d \vec{\lambda} \  |\mu|^2 e^{-\frac{\lambda^2}{2}} \phi_{E=0}(z')^* \phi_{E=0}(z) \exp(\frac{|\mu|^2}{\sqrt{9\pi^2\Lambda}}\times \log z_{IR} ) \label{eq:result_densitymat},
\end{eqnarray}
which has an infinitely strong peak at $\Lambda=0$.
Then, $\Lambda$-integration can be performed simply by substituting $\Lambda=0$ in the integrand, and the exponent in the density matrix can be written as
\beq
 \int^{z_{IR}}_{\epsilon} {dz}\ \frac{1}{z k(z)} \sim  \int^{z_{IR}} {dz}  \frac{1}{z \sqrt{M_{matt}/{z}}} \sim \frac{1}{\sqrt{M_{matt}}}z_{IR}^{1/2},\label{eq:torus_enhancement}
 \eeq
where we have assumed that the universe becomes matter dominated for large $z$. 

\subsection{Open Universes}
For $K=-1$, the potential $U(z)$ is given by 
\begin{equation}
2U(z)=-k_{E=0}^2(z)=-9\Lambda
-\frac{9^{1/3}}{z^{2/3}}
-\frac{2M_{matt}}{z}-\frac{2S_{rad}}{z^{4/3}},\label{eq:open_pot}
\end{equation}
where the second term comes from the negative curvature. 
As in the case of the flat universe, this potential is always negative for $\Lambda >0$, while for $\Lambda<0$ it becomes positive for large $z$.

If all the universes are with $K=-1$,  the density matrix again has a strong peak at $\Lambda=0$.
The exponent in the density matrix then becomes
 \beq
 \int^{z_{IR}}_{\epsilon} {dz}\ \frac{1}{z k(z)} \sim  \int^{z_{IR}} {dz}  \frac{1}{z \sqrt{{z^{-2/3}}}} \sim z_{IR}^{1/3}.\label{eq:open_top}
 \eeq
 \subsection{Summing over topologies}
So far, we have considered the cases that all universes have the same topology. However, we can consider a situation where universes with various topologies appear in the multiverse. In such case, we should sum over topologies in the multiverse wave function. 

To sum over topologies, it is convenient to denote the pair $(z_i,\alpha_i)$, the size and the topology of the $i$-th universe, collectively by $\zeta_i$. Since the probability amplitude $\mu$ may also depend on the topology of the universe, we denote that with topology $\alpha_i$ by $\mu_{\alpha_i}$, or $\mu_{\zeta_i}$.  Then, for the multiverse wave function with various topologies, \eqref{eq:multiverse_state_single} is generalized to 
\begin{eqnarray}
\Phi_N(\zeta_1,\cdots, \zeta_N)= 
\int  d \vec{\lambda} \ \biggl(\prod_{i=1}^N \mu(\zeta_i) \biggr) \times \phi_{E=0}(\zeta_1)\phi_{E=0}(\zeta_2)\cdots\phi_{E=0}(\zeta_{N})\ w(\vec{\lambda}) |  \vec{\lambda} \ket \label{eq:multiverse_state_many}.
\end{eqnarray}
We  compute the density matrix of our universe from this multiverse wave function.
 By introducing a notation
\begin{eqnarray}
\int d \zeta &\equiv& \sum_\alpha \int dz,
\end{eqnarray}
it is given by 
\begin{align}
\rho(\zeta',\zeta) &=  \sum_{{N =0}}^\infty  \int \frac{d \zeta^{N}}{N ! } 
\Phi_{N+1}^* (\zeta',\zeta_1,\cdots,\zeta_N)\Phi_{N+1}(\zeta,\zeta_1,\cdots,\zeta_N)\nonumber \\
&= \sum_{{N =0}}^\infty  \frac{1}{N ! } \int_{-\infty}^{\infty}  d \vec{\lambda}  \ w(\vec{\lambda})^2  \mu^*_{\zeta'} \mu_\zeta\phi_{E=0}(\zeta')^* \phi_{E=0}(\zeta)  \times \biggl( \int d \zeta^{''} |\mu_{\zeta''} \phi_{E=0}(\zeta^{''})|^2 \biggr)^{N} \nonumber \\
\propto& \int_{-\infty}^{\infty}  d \vec{\lambda}  w(\vec{\lambda})^2 \mu^*_{\zeta'} \mu_\zeta\phi_{E=0}(\zeta')^* \phi_{E=0}(\zeta)\exp\biggl( \sum_{\alpha''}\int dz^{''} |\mu_{\alpha''} \phi_{E=0}(z^{''}; K_\alpha)|^2 \biggr).  \label{eq:dens_gen}
\end{align}
We note that, compared with the single topology case \eqref{eq:densitymat}, the exponent becomes the sum over various topologies.

By comparing \eqref{eq:maximum_radiation}, \eqref{eq:torus_enhancement} and \eqref{eq:open_top}, we find that the flat universes dominate in \eqref{eq:dens_gen}.
Therefore, if universes with any topologies are allowed to emerge, the big fix occurs in such a way that $M_{matt}$ in the asymptotic universe with $K=0$ is minimized.
In this case the cosmological constant problem is again solved, but the situation for the other coupling constants differs much from the case of $S_3$ universe.
At this stage we can not tell which case is more realistic, because we have not specified the details about the microscopic dynamics of how  universes emerge from nothing with a small size $z=\epsilon$. 

\section{Summary and Discussion} \label{sec:discussion}
In this paper, we have studied the effect of wormholes on the wave function of the  multiverse and the density matrix of our universe. 
The wormholes make the multiverse wave function a superposition of states with various coupling constants $\{ \lambda_i \}$. 
We have shown that by examining the density matrix $\{ \lambda_i \}$ are determined in such a way that they make the following factor as large as possible 
\beq
\int dz|\phi_{E=0}(z)|^2,  \tag{\ref{eq:time_IR}}
\eeq
which is interpreted as  the lifetime of the universe.
 In particular, it is predicted that the cosmological constant becomes very close to zero in the far future. 
  If we believe the presently observed value of the cosmological constant, which is a non-zero positive value, then our analysis suggests that the cosmological ``constant" will move towards zero such as in the quintessence scenario, where the cosmological constant is the energy of a scalar field rolling down in a runaway potential.	
 
For $S^3$ universes, the coupling constants are determined in such a way that they maximize the lifetime of the universe \eqref{eq:time_IR}. However, it is difficult to search the maximum point of \eqref{eq:time_IR} in the parameter space of $\{ \lambda_i \}$  because \eqref{eq:time_IR} highly depends on which parameters are induced by the wormhole effect and also depends on the cosmology and the physics beyond the standard model such as the dark matter and inflation. As an illustration of the big fix, we made some assumptions on cosmology and studied the possible solution of  the gauge hierarchy problem and the strong CP problem. In particular, our study suggests that the Higgs mass may be fixed at 
 \beq
 m_h\sim 140 \pm 20 \rm{GeV}. \tag{\ref{eq:higgs_pred}}
 \eeq
 
 Although we have mainly studied $S^3$ universe in this paper, there is a possibility that universes are allowed to have  the other topologies as in Section \ref{sec:others}. We found that  in such a situation  our density matrix is determined only from flat universes, and also found that  $\{ \lambda_i \}$ are determined such that $M_{matt}$ in the far future becomes minimized. This naively seems to predict an empty universe and contradict with our universe. Therefore, if the universes are allowed to emerge from nothing with any topologies, there might be some reason in the quantum gravity  that forbids such empty universe to emerge as the initial condition. 

In conclusion, the wormhole mechanism is a fascinating scenario since it can solve naturalness problems in the standard model and the current cosmology without introducing new physics such as supersymmetry or extra dimensions. Although we only have presented an illustration of the big fix scenario, it is interesting to explore the precise prediction  further, and,  for this purpose, the deeper understanding of the quantum gravity is  indispensable.
 

\section*{Acknowledgement} 
The authors acknowledge useful conversations with T. Kobayashi. H.K also thanks Henry Tye for fruitful discussions. This work is supported by the Grant-in-Aid for the Global COE program "The Next Generation of Physics, Spun from Universality and Emergence" from the MEXT.

\appendix
\def\thesection{Appendix \Alph{section}}
\section{ Normalization of the Wave Function  }\label{app:norm}
\def\thesection{ \Alph{section}}

In this appendix, we check the wave function (\ref{eq:general_solution}) satisfies the normalization (\ref{eq:normalization}),
\begin{eqnarray*}
\int^\infty_0 dz \phi^*_{E'} (z) \phi_E (z) = \delta(E-E').
\end{eqnarray*}
Substituting the wave function, the left hand side is
\begin{eqnarray}
\int^\infty_0 dz \frac{1}{\pi z \sqrt{k_E(z)k_{E'}(z)}}\exp( \pm i \int^z dz' ( k_{E'}(z')-k_{E}(z'))). \label{eq:norm_appendix}
\end{eqnarray}
Note that the delta function can arise from the integral over the asymptotic region $z \rightarrow \infty$.
For large $z$, $k_E(z)k_{E'}(z)\sim {9\Lam}$ and $k_{E'}-k_{E}\simeq\frac{\partial k_E}{\partial E}(E'-E)\sim\frac{1}{\sqrt{9\Lam}z}(E'-E)$, where we have used $k^2_E \sim 9\Lam +\frac{2E}{z}+\cdots$.
 From these, we can check (\ref{eq:norm_appendix}) indeed gives,
\begin{eqnarray*}
\int^\infty d (\log z)\frac{1}{\pi \sqrt{9\Lam}}\exp({\pm i \frac{1}{\sqrt{9\Lam}}(E'-E)\log z })=\delta(E'-E).
\end{eqnarray*}

\def\thesection{Appendix \Alph{section}}
\section{The relation between the curvature and e-foldings}\label{app:e-folds}

In this appendix, we relate the e-foldings of the initial inflation to the time when the curvature term becomes comparable to the energy density. 
In section \ref{sec:bigfix}, we have studied the specific case that the curvature term becomes important while the decay products of protons are relativistic.  We will find that this case  corresponds to  the e-foldings given by \eqref{eq:e-folds}.


We denote by $a_*$ the scale factor of the universe when $\Delta S_{rad}$ balances with the curvature. From \eqref{eq:radiation_dominant} and \eqref{eq:decay_conservation}, $a_*$ is given by 
\beq
a_* \simeq \Delta S_{rad}^{1/2}=(a_p  \Delta M_{matt})^{3/2}, \label{eq:astar}
\eeq
where $\Delta M_{matt}$ is the total mass of protons in the whole of the universe.
It can be expressed using the current values of the scale factor $a_0$ and the energy density of protons $\rho_{proton} \simeq 1 \rm{GeV/m^3}$:
\beq
\Delta M_{matt} = a_0^3 \rho_{proton}.\label{eq:deltam}
\eeq

When the scale factor is around $a_p$, the protons decay, and the decay products, especially electrons, are relativistic at that time. However, as the universe expands, the energy of these relativistic electrons scales as  $E_{electron} \propto 1/a$. And when the scale factor becomes about $10^3$ times as large as $a_p$, they will become non-relativistic.\footnote{The number $10^3$ comes from a rough estimate of the ratio between the electron mass and its energy when it is produced by the proton decay.}  
However, from $Condition 3$ in Section \ref{sec:bigfix}, the curvature term must become comparable to the energy density before it happens. Thus, we have the following constraint
 on $a_*$,
\beq
a_p \lesssim a_*  \lesssim a_p \times 10^3.
\eeq 
Substituting \eqref{eq:astar} and \eqref{eq:deltam} into the above equation, we obtain 
\beq
\sqrt{\frac{a_p}{a_0}}\frac{1}{\sqrt{\rho_{proton}}} \lesssim a_0 \lesssim \sqrt{\frac{a_p}{a_0}}\frac{1}{\sqrt{\rho_{proton}}} \times 10^{3}.\label{eq:hutousiki}
\eeq

Next, we estimate the ratio $a_p/a_0$. Since we have assumed that the cosmological constant $\Lambda$ is decreasing from the current value to the asymptotic value $\Lambda_{cr} \simeq 0$, the secondary inflation, which is currently going on, ends within a finite time. We denote  the e-folding during this inflation by $\tilde{N}$. After $\Lambda$ gets sufficiently small and the inflation ends, the protons dominate the energy density, and the universe scales as $a\propto t^{2/3}$. Thus, $a_p/a_0$ is given by 
\beq
a_p/a_0 \sim e^{\tilde{N}}10^{(36-10)\times \frac{2}{3}},\label{eq:protonscale}
\eeq
where we have estimated the proton lifetime as $\tau_p \sim 10^{36} \rm{yr}$ and the age of the universe today as $10^{10} \rm{yr}$. Using \eqref{eq:protonscale} and $\rho_{proton}^{-1/2}\simeq 10^{11}{\rm{ly}}$,\footnote{We are using the unit $G=3\pi/2$.} \eqref{eq:hutousiki} becomes
\beq
e^{\tilde{N}/2} \times 10^{26/3}\lesssim\frac{a_0}{10^{11}{\rm{ly}}} \lesssim e^{\tilde{N}/2} \times 10^{26/3+3},
\eeq
where  $10^{11}{\rm{ly}}$ is the same order as the size of the observable universe and corresponds to  the lower bound on the e-foldings of the initial inflation, $N_{e-fold}>55$. Thus, the above inequality implies\footnote{$10^{26/3}\simeq e^{20}$ and $10^5 \simeq e^{7}$.}
\beq
\frac{\tilde{N}}{2} + 75 \lesssim N_{e-fold} \lesssim \frac{\tilde{N}}{2}+82.\label{eq:e-folds}
\eeq 
Therefore, if $N_{e-fold}$ is in this range, the cosmological assumption we made in Section \ref{sec:bigfix} is  realized.

\bibliographystyle{unsrt}
\bibliography{bunken}

\begin{thebibliography}{10}

\bibitem{wormhole}
\text{For reviews, see, for example, S. Weinberg. The cosmological constant
  problem.} {\it{rev, mod.phys, 61(1), 1989}}.

\bibitem{coleman1988there}
S.~Coleman.
\newblock {Why there is nothing rather than something: A theory of the
  cosmological constant* 1}.
\newblock {\em Nuclear Physics B}, 310(3-4):643--668, 1988.

\bibitem{banks1988prolegomena}
T.~Banks.
\newblock {Prolegomena to a theory of bifurcating universes: a nonlocal
  solution to the cosmological constant problem or little lambda goes back to
  the future}.
\newblock {\em Nuclear Physics B}, 309(3):493--512, 1988.

\bibitem{klebanov1989wormholes}
I.~Klebanov, L.~Susskind, and T.~Banks.
\newblock {Wormholes and the cosmological constant* 1}.
\newblock {\em Nuclear Physics B}, 317(3):665--692, 1989.

\bibitem{polchinski1989phase}
J.~Polchinski.
\newblock {The phase of the sum over spheres}.
\newblock {\em Physics Letters B}, 219(2-3):251--257, 1989.

\bibitem{giddings1989baby}
S.B. Giddings and A.~Strominger.
\newblock {Baby universe, third quantization and the cosmological constant}.
\newblock {\em Nuclear Physics B}, 321(2):481--508, 1989.

\bibitem{PhysRevLett.62.1429}
H.~B. Nielsen and Masao Ninomiya.
\newblock Solution of the strong \textit{CP} problem in baby-universe theory.
\newblock {\em Phys. Rev. Lett.}, 62:1429--1432, Mar 1989.

\bibitem{grinstein1988light}
B.~Grinstein and Mark~B .Wise.
\newblock Light scalars in quantum gravity* 1.
\newblock {\em Physics Letters B}, 212(4):407--410, 1988.

\bibitem{preskill1989wormholes}
J.~Preskill, S.P. Trivedi, and M.B. Wise.
\newblock Wormholes in spacetime and [theta] qcd* 1.
\newblock {\em Physics Letters B}, 223(1):26--31, 1989.

\bibitem{cline1989can}
JM~Cline.
\newblock Can $\theta_{QCD}=\pi$?
\newblock {\em Physical Review Letters;}, 63(13), 1989.

\bibitem{rubakov1988third}
V.A. Rubakov.
\newblock On third quantization and the cosmological constant.
\newblock {\em Physics Letters B}, 214(4):503--507, 1988.

\bibitem{strominger1989lorentzian}
A.~Strominger.
\newblock {A Lorentzian analysis of the cosmological constant problem}.
\newblock {\em Nuclear Physics B}, 319(3):722--732, 1989.

\bibitem{fischler1989quantum}
W.~Fischler, I.~Klebanov, J.~Polchinski, and L.~Susskind.
\newblock Quantum mechanics of the googolplexus.
\newblock {\em Nuclear physics. B}, 327(1):157--177, 1989.

\bibitem{cline1989does}
J.M. Cline.
\newblock {Does the wormhole mechanism for vanishing cosmological constant work
  in lorentzian gravity?}
\newblock {\em Physics Letters B}, 224(1-2):53--57, 1989.

\bibitem{vilenkin1984quantum}
A.~Vilenkin.
\newblock {Quantum creation of universes}.
\newblock {\em Phys. Rev. D;}, 30(2):509--511, 1984.

\bibitem{kawai2011asymptotically}
H.~Kawai and T.~Okada.
\newblock Asymptotically vanishing cosmological constant in the multiverse.
\newblock {\em arXive:1104.1764}, 2011.

\bibitem{Ambjorn:2005qt}
J.~Ambjorn, J.~Jurkiewicz, and R.~Loll.
\newblock {Reconstructing the universe}.
\newblock {\em Phys. Rev.}, D72:064014, 2005.

\bibitem{holland2005triviality}
K.~Holland.
\newblock Triviality and the higgs mass lower bound.
\newblock {\em Nuclear Physics B-Proceedings Supplements}, 140:155--161, 2005.

\bibitem{ishibashi1997large}
N.~Ishibashi, H.~Kawai, Y.~Kitazawa, and A.~Tsuchiya.
\newblock A large-n reduced model as superstring.
\newblock {\em Nuclear Physics B}, 498(1-2):467--491, 1997.

\bibitem{steinacker2007emergent}
H.~Steinacker.
\newblock Emergent gravity from noncommutative gauge theory.
\newblock {\em Journal of High Energy Physics}, 2007:049, 2007.

\bibitem{Kim:2011cr}
Sang-Woo Kim, Jun Nishimura, and Asato Tsuchiya.
\newblock {Expanding (3+1)-dimensional universe from a Lorentzian matrix model
  for superstring theory in (9+1)-dimensions}.
\newblock 2011.
\newblock * Temporary entry *.

\bibitem{gibbons1978path}
G.W. Gibbons, S.W. Hawking, and M.J. Perry.
\newblock Path integrals and the indefiniteness of the gravitational action.
\newblock {\em Nuclear Physics B}, 138(1):141--150, 1978.

\bibitem{hawking1986operator}
SW~Hawking and D.N. Page.
\newblock Operator ordering and the flatness of the universe.
\newblock {\em Nuclear Physics B}, 264:185--196, 1986.

\bibitem{moss1984wave}
IG~Moss and WA~Wright.
\newblock Wave function of the inflationary universe.
\newblock {\em Physical Review D}, 29(6):1067, 1984.

\bibitem{unruh1989time}
W.G. Unruh and R.M. Wald.
\newblock Time and the interpretation of canonical quantum gravity.
\newblock {\em Physical Review D}, 40(8):2598, 1989.

\end{thebibliography}

\end{document}